\documentclass[reprint,aps,pra,showpacs,superscriptaddress,preprintnumbers,amsmath,amssymb,floatfix]{revtex4-1}

\usepackage[pdftex]{graphicx}
\usepackage{dcolumn}
\usepackage{bm}
\usepackage{xcolor}
\usepackage{hyperref}

\newcommand{\Vmat} {\ensuremath{\bm{\mathcal{V}}}}  
\newcommand{\Kmat} {\ensuremath{\bm{\mathcal{K}}}}  
\newcommand{\Cmat} {\ensuremath{\bm{\mathcal{C}}}}  
\newcommand{\Smat} {\ensuremath{\bm{\mathcal{S}}}}  
\newcommand{\Umat} {\ensuremath{\mathbf{U}}}
\newcommand{\etamat} {\ensuremath{\bm{\eta}}}  
\newcommand{\numat} {\ensuremath{\bm{\nu}}}  
\newcommand{\SSmat} {\ensuremath{\mathbf{S}}}
\newcommand{\Xmat} {\ensuremath{\mathbf{X}}}

\graphicspath{{./}}  

\begin{document}

\preprint{To appear in Phys. Rev. A (version as accepted)}   

\title{Assignment of resonances in dissociative recombination of HD$^{+}$ ions: high-resolution measurements compared with accurate
  computations}
 
\author{F.~O.~Waffeu~Tamo} 
\affiliation{Laboratoire Ondes et Milieux Complexes FRE-3102 CNRS and Universit\'e du Havre, 25, rue Philippe Lebon, BP 540, 76058, Le
  Havre, France}
\affiliation{Centre for Atomic, Molecular Physics and Quantum Optics (CEPAMOQ), University of Douala, P. O. Box 8580, Douala, Cameroon}
\affiliation{Laboratoire d'Etude du Rayonnement et de la Mati\`ere en Astrophysique, Observatoire de Paris, 91295 Meudon Cedex, France}

\author{H.~Buhr}
\affiliation{Max-Planck-Institut f\"ur Kernphysik, Saupfercheckweg 1, 69117 Heidelberg, Germany}
\affiliation{Department of Particle Physics, Weizmann Institute of Science, P. O. Box 26, 76100 Rehovot, Israel}

\author{O.~Motapon}
\affiliation{LPF, UFD Physique et Sciences de l'Ing\'enieur, University of Douala, P. O. Box 24157, Douala, Cameroon}

\author{S.~Altevogt}
\affiliation{Max-Planck-Institut f\"ur Kernphysik, Saupfercheckweg 1, 69117 Heidelberg, Germany}

\author{V.~M.~Andrianarijaona}
\altaffiliation[Present address: ]{Department of Physics, Pacific Union College, Angwin, CA 94508, USA}
\affiliation{Max-Planck-Institut f\"ur Kernphysik, Saupfercheckweg 1, 69117 Heidelberg, Germany}

\author{M.~Grieser}
\affiliation{Max-Planck-Institut f\"ur Kernphysik, Saupfercheckweg 1, 69117 Heidelberg, Germany}

\author{L.~Lammich}
\altaffiliation[Present address: ]{Institute of Physics and Astronomy, University of Aarhus, 8000 Aarhus C, Denmark}
\affiliation{Max-Planck-Institut f\"ur Kernphysik, Saupfercheckweg 1, 69117 Heidelberg, Germany}

\author{M.~Lestinsky}
\altaffiliation[Present address: ]{GSI Helmholtzzentrum f\"ur Schwerionenfor\-schung, 64291 Darmstadt, Germany}
\affiliation{Max-Planck-Institut f\"ur Kernphysik, Saupfercheckweg 1, 69117 Heidelberg, Germany}

\author{M.~Motsch}
\altaffiliation[Present address: ]{Laboratorium f\"ur Physikalische Chemie, ETH Z\"urich, 8093 Zurich, Switzerland}
\affiliation{Max-Planck-Institut f\"ur Kernphysik, Saupfercheckweg 1, 69117 Heidelberg, Germany}

\author{I.~Nevo}
\affiliation{Department of Particle Physics, Weizmann Institute of Science, P. O. Box 26, 76100 Rehovot, Israel}

\author{S.~Novotny}
\affiliation{Max-Planck-Institut f\"ur Kernphysik, Saupfercheckweg 1, 69117 Heidelberg, Germany}

\author{D.~A.~Orlov}
\affiliation{Max-Planck-Institut f\"ur Kernphysik, Saupfercheckweg 1, 69117 Heidelberg, Germany}

\author{H.~B.~Pedersen}
\altaffiliation[Present address: ]{Institute of Physics and Astronomy, University of Aarhus, 8000 Aarhus C, Denmark}
\affiliation{Max-Planck-Institut f\"ur Kernphysik, Saupfercheckweg 1, 69117 Heidelberg, Germany}

\author{D.~Schwalm}
\affiliation{Max-Planck-Institut f\"ur Kernphysik, Saupfercheckweg 1, 69117 Heidelberg, Germany}
\affiliation{Department of Particle Physics, Weizmann Institute of Science, P. O. Box 26, 76100 Rehovot, Israel}

\author{F.~Sprenger}
\affiliation{Max-Planck-Institut f\"ur Kernphysik, Saupfercheckweg 1, 69117 Heidelberg, Germany}

\author{X.~Urbain}
\affiliation{Institute of Condensed Matter and Nanosciences, Universit\'e catholique de Louvain, chemin du cyclotron 2, 1348 Louvain-la-Neuve, Belgium}

\author{U.~Weigel} 
\affiliation{Max-Planck-Institut f\"ur Kernphysik, Saupfercheckweg 1, 69117 Heidelberg, Germany}

\author{A.~Wolf}
\email[Email: ]{A.Wolf@mpi-hd.mpg.de}
\affiliation{Max-Planck-Institut f\"ur Kernphysik, Saupfercheckweg 1, 69117 Heidelberg, Germany}

\author{I.~F.~Schneider}
\altaffiliation[Work partly performed in visit at ]{Laboratoire Aim{\'e}--Cotton, Universit\'e Paris-Sud, 91405 Orsay Cedex, France}
\email[Email: ]{ioan.schneider@univ-lehavre.fr}
\affiliation{Laboratoire Ondes et Milieux Complexes FRE-3102 CNRS and Universit\'e du Havre, 25, rue Philippe Lebon, BP 540, 76058, Le Havre, France}

\date{\today}

\begin{abstract}
  The collision-energy resolved rate coefficient for dissociative recombination of HD$^+$ ions in the vibrational ground state is
  measured using the photocathode electron target at the heavy-ion storage ring TSR.  Rydberg resonances associated with ro-vibrational
  excitation of the HD$^+$ core are scanned as a function of the electron collision energy with an instrumental broadening below 1 meV
  in the low-energy limit.  The measurement is compared to calculations using multichannel quantum defect theory, accounting for
  rotational structure and interactions and considering the six lowest rotational energy levels as initial ionic states.  Using thermal
  equilibrium level populations at 300 K to approximate the experimental conditions, close correspondence between calculated and
  measured structures is found up to the first vibrational excitation threshold of the cations near 0.24 eV.  Detailed assignments,
  including naturally broadened and overlapping Rydberg resonances, are performed for all structures up to 0.024 eV.  Resonances from
  purely rotational excitation of the ion core are found to have similar strengths as those involving vibrational excitation.  A
  dominant low-energy resonance is assigned to contributions from excited rotational states only.  The results indicate strong
  modifications in the energy dependence of the dissociative recombination rate coefficient through the rotational excitation of the
  parent ions, and underline the need for studies with rotationally cold species to obtain results reflecting low-temperature ionized
  media.
\end{abstract}

\maketitle

\section{Introduction }

The dissociative recombination (DR) process \cite{drbook} of diatomic ions, together with its competing reactions -- ro-vibrational
excitation and de-excitation as well as, at higher energy, dissociative excitation -- plays a decisive role in astrophysical
\cite{petrie-msr07} and planetary \cite{witasse08} ionized media, fusion plasma in the divertor region \cite{krasheninnikov-physscr-02}
and in most cold plasmas of technological interest \cite{goodings-combflame07}.  The hydrogen cation and its isotopologues represent
benchmark systems frequently taken up in experimental and theoretical studies.  With the technique of merged electron and ion beams
\cite{phaneuf-rpp99}, collision studies can be realized under quasi-monoenergetic conditions for detailed comparisons of experiment and
theory.  This yields collision-energy resolved rate coefficients while, for sensitive comparison to theory, good control is needed over
the ro-vibrational excitation of the incident cations.  Following first results on H$_{2}^+$ in single-pass merged beams \cite{hus88},
the infrared active isotopologue HD$^+$ moved into the focus of experiments \cite{forck93,tanabe95,stromholm95,alkhalili03} using the
ion-storage ring technique \cite{larsson-rpp95}, where fast ion beams are stored over many seconds and vibrational de-excitation of
HD$^+$ is completed within much less than a second \cite{amitay98}.  On H$_{2}^+$, storage-ring experiments gave access to vibrational
de-excitation in electron collisions \cite{krohn00,mot08}.  Moreover, using special ion sources, the control of the initial states
could be much improved in recent experiments \cite{zhaunerchyk07,deruette07,novotny-thesis08} on H$_{2}^+$ + $e^-$.  They resulted in
accurate state-to-state, energy-resolved rate coefficients whose comparison to advanced theoretical studies \cite{takagi02,mot08} is
underway.

Studies on the system HD$^+$ + $e^-$ have reached a high level of accuracy.  Storage-ring experiments were compared to detailed
predictions \cite{ifs-a18,takagi09} from multichannel quantum defect theory (MQDT) that addressed in particular the dependence of the
rate coefficient on the rotational excitation of the ion.  The energy-resolved DR rate coefficient \cite{alkhalili03} and rotational
effects in the resonant collision processes \cite{novotny08:prl:anisotropy} were experimentally investigated.  Moreover, experimental
and theoretical results were compared for the related process of rotational de-excitation in low-energy electron collisions
\cite{shafir09}.

The study presented here on the DR reaction
\begin{equation}
  \label{eq:DR} {\text H}{\text D}^{+} + e^{-} \rightarrow
\left\lbrace\begin{array}[c]{c} ~{\text H} + {\text D}^*\\[2mm]{\text H}^{*} + {\text D}\end{array}\right. ,
\end{equation}
is motivated by the large improvement of the experimental energy resolution achieved by using a photocathode electron beam
\cite{orlov04} for merged-beams experiments \cite{aw-jpcs09}, in which electron temperatures below 1\,meV are reached.  As a function
of the collision energy, scanned with high resolution, the DR rate coefficient displays a rich structure of resonances, caused by
resonant capture of the incident electron into Rydberg states on a ro-vibrationally excited HD$^+$ core before the dissociation of the
recombined system into the channels given by Eq.\ (\ref{eq:DR}).  These measurements will be presented together with a detailed
comparison to state-by-state theoretical calculations of the DR rate coefficient by the MQDT method with the goal to assign the
dominant resonance structures.  The assignment provides a highly sensitive test for the quality of the molecular data -- potential
energy surfaces and mutual couplings -- and for the theoretical approach to account for the underlying mechanisms and interactions.

The setup allowing our current measurements is described in Sec.~\ref{sec_expsetup}.  Section~\ref{sec_expresults} presents our
experimental results.  The main steps in the MQDT computation of the cross section are reviewed in Sec.~\ref{sec_mqdt}, and the
procedure of resonance assignment is described in Sec.~\ref{sec_assignment}.  A detailed comparison between measurements and
computations is performed in Sec.~\ref{sec_comparison}.

\section{\label{sec_expsetup} Experimental setup}

The experiment was performed in the heavy ion storage ring TSR at the Max Planck Institute for Nuclear Physics in Heidelberg, Germany.
HD$^+$ ions from a Penning source were accelerated in a radiofrequency quadrupole accelerator and a linear rf accelerator to a final
energy 5.3~MeV, injected into the storage ring within 30\,$\mu$s, and stored for up to 10\,s before their replacement by a new
injection.  Along their closed orbit with a circumference of 55.4\,m, the residual gas pressure is below 1$\times$10$^{-10}$~mbar.  As
in previous experiments \cite{lestinsky-prl08} at this facility, the ions were merged on each turn in the ring with collinear,
velocity-matched or slightly velocity-detuned electron beams of $\sim$1.5\,m length in two straight sections of the TSR: the electron
cooler \cite{steck-nima90,pastuszka-nima96,schmitt-thesis99}, and the electron target section (ETS) \cite{sprenger04}.

The electron cooler was used to compress the phase-space density of the injected ion beam within the first 1--2\,s after the injection,
reaching $\sim$1\,mm ion beam diameter, much smaller than the electron beam diameters of $\sim$27\,mm and $\sim$19\,mm in the electron
cooler and in the ETS, respectively.  The electron cooler with beam velocity $v_{e,c}$ locked the ion beam velocity to $v_i=v_{e,c}$;
the laboratory electron energy of this electron beam was near 970\,eV.  It was generated by a thermionic cathode and magnetically
expanded so that the temperature in the co-moving frame of the beam can be calculated \cite{pastuszka-nima96,schmitt-thesis99} to
$k_BT_\perp=16$\,meV.  This temperature refers to the transverse direction perpendicular to the ion beam, while the longitudinal
electron velocity differences correspond to a temperature found to be $k_BT_\|\sim0.06$\,meV in independent measurements
\cite{gwinner-prl00}.  The electron density of the electron cooler amounted to $n_{e,c}$=1.8$\times$10$^7$~cm$^{-3}$ and a longitudinal
magnetic field of 0.04\,T was present in its interaction region with the ion beam.

The electron beam in the ETS was used for the DR rate coefficient measurements at variable beam velocity $v_{e,t}$, which sets the
electron--ion collision energy through the detuning energy defined as
\begin{equation}
\label{E_d}
E_d = (m_e/2) \left(v_i - v_{e,t}\right)^2 ,
\end{equation}
where $m_e$ is the electron mass.  The electron--ion collision velocities are described by an anisotropic thermal distribution
\cite{andersen-pra90} with the transverse and longitudinal electron temperatures $T_\perp$ and $T_\|$, respectively.  The electron beam
of the ETS was produced with a photocathode at liquid-nitrogen temperature~\cite{orlov04} and, similar to the electron cooler, magnetic
expansion was applied between the cathode and the interaction region \cite{sprenger04}.  At matched beam velocities
($v_{e,t}=v_i=v_{e,c}$, corresponding to $E_d=0$) the electron density of the ETS was $n_{e,t}=3.2\times10^5$~cm$^{-3}$; a longitudinal
magnetic field of 0.057\,T was present in its interaction region.  Electron temperatures of $k_BT_\perp=(0.6\pm0.2)$\,meV and $k_BT_\|$
near 25\,$\mu$eV were determined for the photocathode electron beam in the present measurement, observing energy dependences in the DR
rate coefficient as described below.

The neutral fragments emerging from the electron--ion reactions in the ETS travel with the ion beam velocity and leave the storage ring
in forward direction towards a dedicated beamline in which a surface barrier detector (SBD) with 48$\times$48~mm$^2$ active area is
situated at about 11.5~m from the ETS center. The SBD has unit efficiency and yields signals proportional to the kinetic energy of the
fragments impinging within the sampling time of 1~$\mu$s which is long compared to the arrival time difference of fragments from one DR
event of $\sim$1~ns, while short compared to the typical interval between two independent events at rates of 1~kHz. Signals associated
with the full mass of HD$^+$ are associated with a DR event, whereas those corresponding to smaller masses are treated as events
yielding a single neutral H or D.

After the ion beam is injected into the TSR, phase-space cooling and radiative cooling of the molecular ions are allowed to proceed
over a period of typically 2\,s before data are taken.  The laboratory energy $E_{e,t}$ of the electrons in the ETS is then varied in
cycles of tens of milliseconds between values of $E_{e,t}=E_c$, which yield $E_d=0$ (``cooling''), of $E_{e,t}=E_r$, which yield
$E_d=10$~eV (``reference''), and of $E_{e,t}=E_m$ (``measurement''), varied according to the range of detuning energy $E_d$ to be
scanned.  The rates of DR signals, $R_{\text{DR}}$, and of single mass events, $R_i$ with $i= \text{H},\text{D}$, are recorded at these
energies.  The variation of $E_m$ from injection to injection yields the rates as a function of energy, from which together with the
electron densities the relative DR rate coefficient at $E_m$ is obtained by
\begin{equation}
  \alpha_{\text{DR}}(E_m) = \frac{R_{\text{DR}}(E_m)}{R_{\text{D}}(E_c,E_m)} \, 
  \frac{\tilde R_{\text{D}}(E_c)}{\tilde R_{\text{DR}}(E_r)} \, \frac{n_{e,t}(E_r)}{n_{e,t}(E_m)}.
\end{equation}
Here, $R_{\text{DR}}(E_m)$ is the DR rate at $E_m$ and $R_{\text{D}}(E_c,E_m)$ the rate of deuterons at $E_c$, both averaged over all
cycles of a given $E_m$, whereas $\tilde R_{\text{D}}(E_c)$ and $\tilde R_{\text{DR}}(E_r)$ are the rates of deuterons and of DR at
$E_c$ and $E_r$, respectively, averaged over all values of $E_m$ of one data set.  Different data sets may differ in the ratio of
${\tilde R_{\text{D}}(E_c)}/{\tilde R_{\text{DR}}(E_r)}$ depending on the average pressure in the storage ring.

As a function of the laboratory energy $E_m$, the rate coefficient $\alpha_{\text{DR}}$ measured by this method shows a peak at
$E_m=E_c$, corresponding to the smallest collision energies $\varepsilon$ where the DR cross section $\sigma_{\text{DR}}(\varepsilon)$
diverges, while the rate coefficient $\alpha_{\text{DR}}(E_m)$ assumes a finite value determined by the spread of the electron--ion
collision velocities.  From Eq.\ (\ref{E_d}), relating $E_{e,t}$ to $E_d$, it is seen that any variations of $E_{e,t}$ lead to much
smaller changes in the collision energies, distributed in a narrow range around $E_d$.  Hence, sorting the data into suitable intervals
of $E_d$, highly resolved variations of the energy resolved merged-beams rate coefficient $\alpha_{\text{DR}}(E_d)$ are measured.  This
rate coefficient is linked to the DR cross section via
\begin{equation}
\label{eq:ratecoefficient}
\alpha_\text{DR}(E_d) =  \int_0^\infty v\sigma_\text{DR}(\varepsilon) \rho(\varepsilon,E_d)\, d\varepsilon
\end{equation}
where $v=(2\varepsilon/m_e)^{1/2}$ is the collision velocity and $\rho(\varepsilon,E_d)$ the distribution of collision energies for a
given detuning energy, which follows from the anisotropic thermal velocity distribution and reflects the electron beam temperatures
$T_\perp$ and $T_\|$.  For the transformation between $E_{e,t}$ and $E_d$, small long-term drifts between the voltages supplied in the
ETS and the electron cooler are monitored by regular scans over the peak at $E_m=E_c$, fixing $E_d=0$ at the maximum position of the
peak in the analysis of the data taken in the temporal vicinity.  The weighted average of these rate coefficients for the different
data sets is then corrected for toroidal effects~\cite{lampert96} and finally absolutely normalized to the rate coefficient
$\alpha_{\text{DR}}(E_d\!=\!\text{9.8~eV})=6.6\times10^{-9}$~cm$^3$\,s$^{-1}$ reported by \textcite{alkhalili03}.

In addition to the energy-resolved velocity-averaged rate coefficient $\alpha_\text{DR}(E_d)$, also the reduced rate coefficient
$\alpha_\text{DR}(E_d)E_d^{1/2}$ is used for the presentation of experimental and theoretical data.  This choice suppresses the overall
decrease of $\alpha_\text{DR}(E_d)$, scaling as $\propto1/E_d^{1/2}$, which is caused by the $1/\varepsilon$ behaviour of the
non-resonant DR cross section.  For a cross section $\sigma_\text{DR}(\varepsilon) = K/\varepsilon$ with constant $K$, this reduced DR
rate coefficient for $E_d\gg k_BT_\perp$ assumes the constant value of $K(2/m_e)^{1/2}$.  Near $E_d=0$, the value of
$\alpha_\text{DR}(E_d)E_d^{1/2}$ is lowered by the convolution over the electron energy distribution.  For asymptotically small $E_d$,
it rises as $K(2\pi E_d/m_ek_BT_\perp)^{1/2}$ because of the convolution with the electron energy distribution and reaches $>$90\% of
its final value at $E_d>4k_BT_\perp$, which corresponds to $\sim$2.5\,meV in the present case.  Thus, above a few meV, variations of
$\alpha_\text{DR}(E_d)E_d^{1/2}$ only occur if the energy dependence of $\sigma_\text{DR}$ deviates from $1/\varepsilon$, which
emphasizes the resonant contributions.

\section{\label{sec_expresults} Experimental results}

\begin{figure}[tb]
  \center\includegraphics[width=8.5cm]{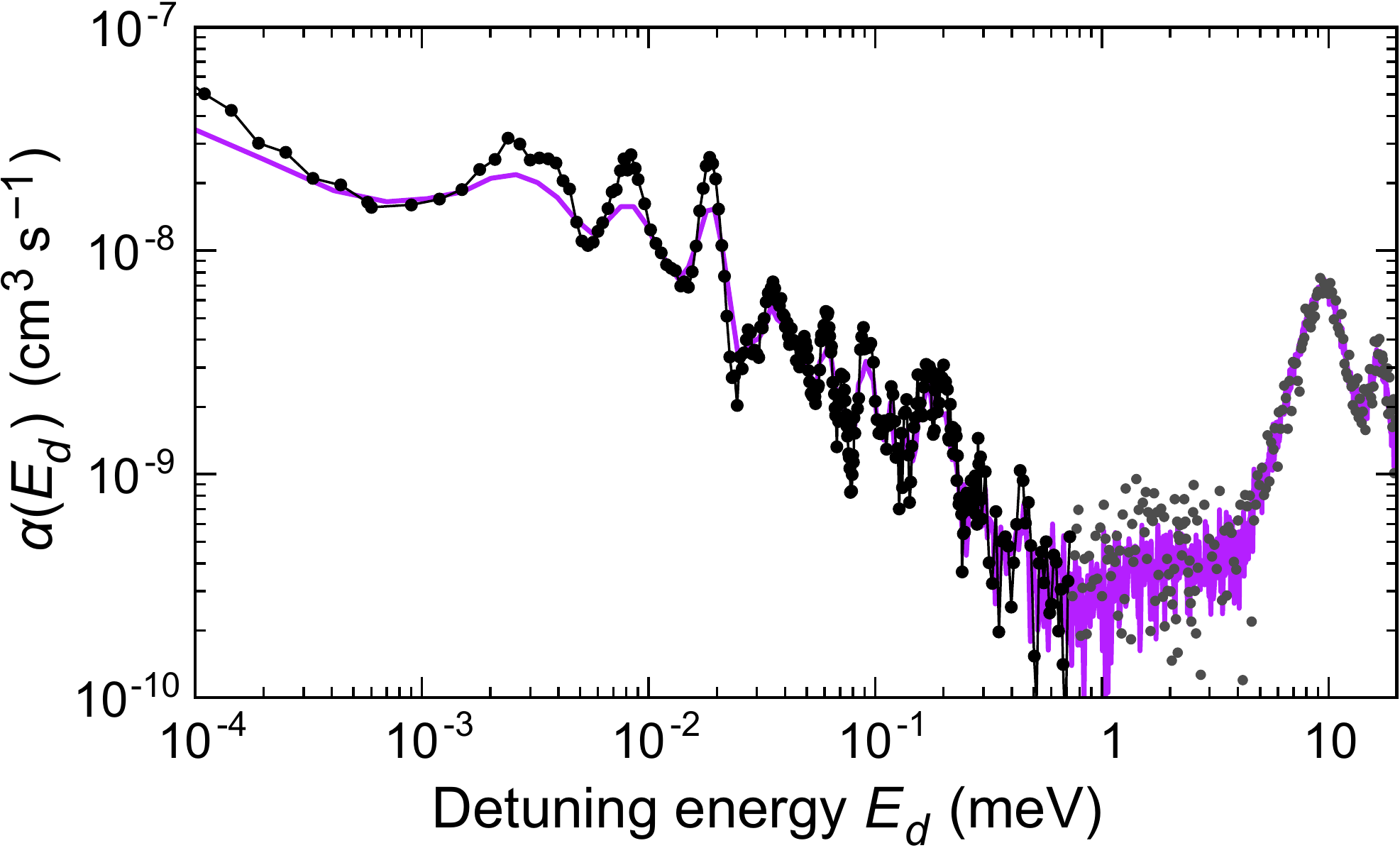}
  \caption{\label{cap:exp_rc} (Color online) DR rate coefficient of HD$^+$ in the range of 10$^{-4}$ to 20~eV, measured with a cold
    electron beam from a photocathode ($kT_\perp$=0.6(2)~meV, black line and symbols). This is compared to the previous benchmark
    measurement~\cite{alkhalili03} ($kT_\perp$=2~meV) shown as a violet (gray) line.  }
\end{figure}

The measured merged-beams rate coefficient as function of the detuning energy is shown in Fig.\ \ref{cap:exp_rc}.  Temporal changes of
the energy dependence became insignificant for times larger than 2\,s after the injection, so that we present data from the storage
time interval of 2--10\,s.  High-resolution scans were limited to the energy region below 0.7\,eV, above which the rate coefficient
becomes very small.  The high-energy peaks near 10 and 16\,eV were scanned with lower resolution and less measurement time.  Here the
previously observed structure \cite{alkhalili03} for vibrationally relaxed HD$^+$ ions is well reproduced.  Of the previous benchmark
measurements presented by \textcite{alkhalili03}, that from CRYRING is chosen for comparison with the present data as it was obtained
with the lowest electron temperature so far reported for this system, $k_BT_\perp=2$\,meV.  In the present results, the structure at
low energies appears much better resolved.  Thus, the peaks at 3, 8, and 19 meV become more pronounced and sub-structures appear much
more clearly, as illustrated in Fig.\ \ref{cap:exp_rc} by the additional low-energy peak near 2.5\,meV.

\begin{figure}[tb]
  \center\includegraphics[width=8.5cm]{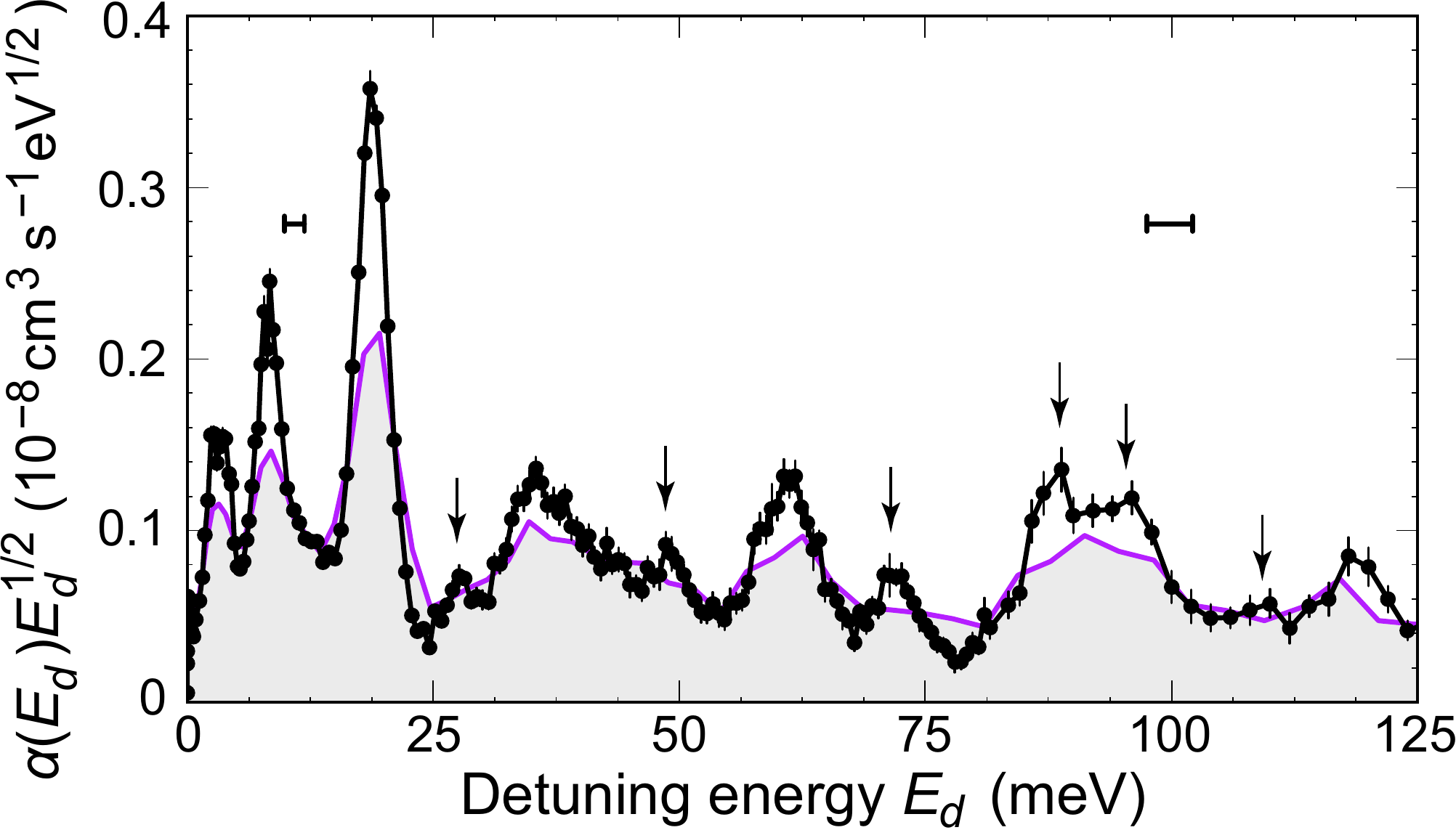}
  \caption{\label{cap:exp_rc_red} (Color online) Reduced DR rate coefficient of HD$^+$ in the low energy range, measured with a cold
    electron beam from a photocathode ($kT_\perp$=0.6(2)~meV, black line and symbols). This is compared to the previous benchmark
    measurement~\cite{alkhalili03} ($kT_\perp$=2~meV) shown as violet (gray) line.  Newly resloved structures are highlighted by
    arrows.  The horizontal bars indicate the collision energy spread (FWHM) of the photocathode measurement in the respective energy
    regions.}
\end{figure}

A more detailed view of the low-energy structure is presented in Fig.\ \ref{cap:exp_rc_red} using the reduced rate coefficient
$\alpha_\text{DR}(E_d)E_d^{1/2}$.  This reveals clear sub-structures and newly observed peaks between 30 and 100\,meV.

For a determination of the electron temperature, the low-energy edges of the lowest-lying peaks, corresponding to energy regions near
2, 7, and 17\,meV, were compared to a model rate coefficient.  It was obtained from a cross section composed of an overall
$1/\varepsilon$ decrease and three infinitely narrow resonances near the peak positions, folded with the anisotropic thermal velocity
distribution.  For appropriately chosen peak positions, good agreement was found for $k_BT_\perp=(0.6\pm0.2)$\,meV, while the
high-energy edges are consistent with $k_BT_\|=0.025$\,meV.  This implies a collision energy spread (full width at half maximum, FWHM)
of 1 meV or less in the low-energy limit, $\sim$2 meV at $E_d=10$\,meV and $\sim$6.5\,meV at $E_d=200$\,meV, scaling as $\propto
E_d^{1/2}$ for $E_d\gtrsim80$\,meV.
%
%
\begin{table}[b]%
  \caption{Properties of rotational levels in the vibrational ground state HD$^+$, showing $N_i^+$ (first line), 
    excitation energies in meV (second line, from Ref.\ \cite{hunter-adndt74}), relative thermal populations at 300\,K 
    (third line), and radiative lifetimes in s (fourth line, from Ref.\ \cite{amitay94}).\label{tab:rot}}%
\begin{ruledtabular}%
\begin{tabular}{cccccccc}%
  0 & 1 & 2 & 3 & 4 & 5 & 6 & $>$6 \\
  \hline\\[-3mm]
  0   &  5.44 &  16.28 &  32.47 &  53.91 &  80.46 &  111.99 \\
  0.104 & 0.251 & 0.271 & 0.199 & 0.108& 0.046 & 0.015 & 0.006\\
  & 140.2 & 14.61 & 4.04 & 1.64 & 0.823 & 0.469 & \\
\end{tabular}
\end{ruledtabular}
\end{table}
The energy resolution obtained in the present measurements is thus comparable to, and for very low collision energies even below the
energy splitting of the two lowest rotational levels of HD$^+$ of 5.44 meV, so that the rotational structure of the ro-vibrationally
excited Rydberg resonances \cite{bardsley68} in the electron collision continuum can be expected to become visible.

After leaving the ion source the HD$^+$ ions are vibrationally and rotationally excited (rotational and vibrational initial ionic
levels $N_i^+$ and $v_i^+$, respectively).  But while the vibrations are known to radiatively relax within storage times considerably
shorter than 1\,s \cite{amitay98}, the rotational excitation of the HD$^+$ ions built on the vibrational ground state is expected to
reach an equilibrium only after several seconds of storage, with a number of rotational states remaining populated at all times.  The
cooling rate and the resulting population distribution finally reached are determined by the interaction with the blackbody radiation
field of 300\,K that pervades the inner volume of the room-temperature ion storage ring, and by the continuous interaction of the
HD$^+$ ions with the electrons of the electron cooler and the electron target.  While the Einstein coefficients between the rotational
states required to describe the interaction of HD$^+$ with the radiation field are well known \cite{amitay94}, the expected large
inelastic cross sections for low-energetic electron scattering on rotational states of HD$^+$ could only recently be verified by
observing the rotational cooling of HD$^+$ by super-elastic electron collisions (SEC) in a dedicated experiment \cite{shafir09}.  Based
on these findings we estimated the time development of the rotational populations by adjusting the calculations described by
\textcite{shafir09} to the present experimental situation.  In particular, the larger temperature of the electron beam of the electron
cooler of 16\,meV in its co-moving reference frame, substantially higher than that of the electron beam used in Ref.\ \cite{shafir09}
(2.8 meV), led not only to rotational de-excitation by SEC.  It also caused considerable excitation of the lowest rotational states by
inelastic collisions with electrons in the high-energetic tail of the velocity distribution in the electron cooler.  Conversely,
cooling and heating of the $N_i^+$ distribution by inelastic electron collisions in the ETS were found to be negligible, mainly because
of its much lower electron density.  While steady-state conditions were found in these estimates to be reached only asymptotically at
the end of the observation window ($\sim$10\,s), the average $N_i^+$ population distribution, weighted by the storage lifetime of the
HD$^+$ beam of $\sim$7\,s, can be approximated by a Boltzmann distribution with a rotational temperature in the range of 250--300\,K
(the range of uncertainty reflecting the limited accuracy at which the SEC cross sections could be verified in Ref.\ \cite{shafir09}).
In the following discussion we therefore assume a thermal distribution of 300\,K to reasonably represent the relative populations of
the HD$^+$ rotational levels during the present measurement.  The resulting populations of the seven lowest rotational states with
quantum numbers $N_i^+ = 0$--6 are compiled together with other properties in Table~\ref{tab:rot}.

\section{\label{sec_mqdt} Theoretical Method}

Dissociative recombination is a reactive collision involving \emph{ionization} channels, describing the scattering of an electron on
the molecular ion, and \emph{dissociation} channels, accounting for the atom--atom scattering.  The main steps of our current MQDT
treatment \cite{mot08} are described below.

\subsection{Construction of the interaction matrix {\boldmath $\mathcal{V}$}}  

This is performed in the outer shell of the region of small electron--ion and nucleus--nucleus distances, that is in the `A-region'
\cite{ja77}, where the Born-Oppenheimer context is appropriate for the description of the collision system.  The good quantum numbers
in this region are $N$, $M$, and $\Lambda$, associated respectively to the total angular momentum and its projections on the $z$-axes
of the laboratory-fixed and of the molecule-fixed frames.

Within a \emph{quasi-diabatic representation} \cite{sidis71,bardsley68,annick80}, the relevant states are organized in \emph{channels},
according to the type of fragmentation which they are meant to describe.
An \emph{ionization} channel is built starting from the ground electronic state of the ion and one of its ro-vibrational levels $N^{+}
v^{+}$, and is completed by gathering all the mono-electronic states of a given orbital quantum number $l$.  These mono-electronic
states describe, \emph{with respect to the $N^{+}v^{+}$ threshold}, either a ``\emph{free}'' electron -- in which case the total state
corresponds to \emph{(auto)ionization} -- or to a \emph{bound} electron -- in which case the total state corresponds to a temporary
\emph{capture into a Rydberg state}.  In the A-region, these states may be modeled reasonably well with respect to the hydrogenic
states in terms of the quantum defect $\mu^{\Lambda}_{l}(R)$, dependent on the internuclear distance $R$, but assumed to be
\emph{independent of energy}.

An ionization channel is coupled to a \emph{dissociation} one, labeled $d_{j}$, by the electrostatic interaction $1/r_{12}$.  In the
molecular-orbital (MO) picture, the states corresponding to the coupled channels must differ by at least two orbitals, the dissociative
states being doubly-excited for the present case.  We account for this coupling at the \emph{electronic} level first, through an
$R$-dependent \emph{scaled} ``Rydberg--valence'' interaction term, $V^{(e)\Lambda}_{d_{j},l}$, assumed to be \emph{independent of the
  energy} of the electronic states pertaining to the ionization channel.  Subsequently, the integration of this electronic interaction
on the internuclear motion results in the elements of the interaction matrix \Vmat,
\begin{equation}
  \label{vmat}
  \mathcal{V}_{d_{j},lN^{+}v^{+}}^{N\Lambda}(E,E)=
  \langle\chi^{\Lambda}_{Nd_{j}}|V^{(e)\Lambda}_{d_{j},l}|\chi^{\Lambda^+}_{N^{+}v^{+}}\rangle,
\end{equation}
where $E$ is the total energy and $\chi^{\Lambda}_{Nd_{j}}$ and $\chi^{\Lambda^+}_{N^{+}v^{+}}$ are the nuclear wave-functions
corresponding to a dissociative state and to an ionization channel, respectively.

This procedure applies in each $\Lambda$-subspace, and results in a block-diagonal global interaction matrix.  The block-diagonal
structure, corresponding to the available $\Lambda$ symmetries, propagates to the further built matrices.

\subsection{Construction of the reaction matrix {\boldmath $\mathcal{K}$}}  

Starting from the interaction matrix \Vmat, we build the reaction K-matrix, which satisfies the Lippmann-Schwinger integral equation
\begin{equation}
  \label{eq:Lippmann-Schwinger}\Kmat = \Vmat + \Vmat \frac{1}{E-\boldsymbol{H}_0} \Kmat.
\end{equation}
This equation has to be solved once \Vmat\ -- whose elements are given by Eq.\ (\ref{vmat}) -- is determined.  Here $\boldsymbol{H}_0$
is the zero order Hamiltonian associated to the molecular system, while neglecting the interaction potential \Vmat.  It has been proven
that, provided the electronic couplings are energy-independent, a perturbative solution of Eq.~(\ref{eq:Lippmann-Schwinger}) is exact
to second order \cite{ngassam03b}.

\vspace{2mm}
\subsection{Diagonalization of the reaction matrix {\boldmath $\mathcal{K}$}}  

In order to express the result of the short-range interaction in terms of phase-shifts, we perform a unitary transformation of our
initial basis into eigenstates.  The columns of the corresponding transformation matrix {\Umat} are the eigenvectors of the K-matrix,
and its eigenvalues serve to evaluate the non-vanishing elements of a diagonal matrix \etamat:
\begin{equation}
  \Kmat \Umat = -{\frac{1}{\pi}}\tan(\etamat)\Umat.
\label{K-pvp}
\end{equation}

\subsection{Frame transformation to the external region}

In the external zone -- the `B-region' \cite{ja77}, characterized by large electron-core distances -- the Born-Oppenheimer model is no
longer valid for the whole molecule, but only for the ionic core.  $\Lambda$ is no longer a good quantum number, and a frame
transformation \cite{fano70,chfan,valcu} is performed between coupling schemes corresponding to the incident electron being decoupled
from the core electrons (external region) or coupled to them (internal region).  The frame transformation coefficients involve angular
coupling, electronic and vibrational factors, and are given by
\begin{widetext}
\begin{eqnarray}
\label{eq:coeffCv}
{\cal C}_{lN^{+}v^{+}, \Lambda \alpha} & = &
\left( \frac{2N^{+}+1}{2N+1}\right)^{1/2}
\left\langle l\left( \Lambda -\Lambda ^{+}\right) N^{+}\Lambda^{+}|lN^{+}N\Lambda \right\rangle\\
& & \times\frac{1+\tau^{+}\tau\left(-1 \right)^{N-l-N^{+}}} {\left[2\left(2-\delta_{\Lambda^{+},0} \right)
  \left(1+ \delta_{\Lambda^{+},0}\delta_{\Lambda,0} \right)   \right] ^{1/2}} 
\times\sum_{v} U_{lv,\alpha}^{\Lambda}\langle
\chi_{N^{+}v^{+}}^{\Lambda ^{+}}| \cos(\pi \mu_{l}^{\Lambda}(R)+\eta_{\alpha}^{\Lambda})|\chi_{Nv}^{\Lambda}\rangle 
\nonumber\\ 
%
\label{eq:coeffCd}{\cal C}_{d_{j},\Lambda \alpha} & = & U_{d_{j}\alpha}^{\Lambda}\cos \eta_{\alpha}^{\Lambda}
\end{eqnarray}
\end{widetext}
as well as by ${\cal S}_{lN^{+}v^{+},\Lambda \alpha }$ and ${\cal S}_{d_{j},\Lambda \alpha }$, which are obtained by replacing cosine
with sine in Eqs.\ (\ref{eq:coeffCv}) and (\ref{eq:coeffCd}).  In the preceding formulae, $\chi_{N^{+}v^{+}}^{\Lambda ^{+}}$ is a
vibrational wavefunction of the molecular ion, and $\chi_{Nv}^\Lambda$ is a vibrational wavefunction of the neutral system adapted to
the interaction (A) region.  The matrix elements $\langle \chi_{N^{+}v^{+}}^{\Lambda ^{+}}| \cos(\pi \mu_{l}^{\Lambda}
(R)+\eta_{\alpha}^{\Lambda})|\chi_{Nv}^{\Lambda}\rangle$ represent the ro-vibronic couplings giving rise to the resonant capture
processes.  The quantities $\tau^{+}$ and $\tau$ are related to the reflection symmetry of the ion and neutral wave function
respectively, and take the values +1 for symmetric states and $-$1 for antisymmetric ones.  The ratio in front of the sum in the right
member of Eq.\ (\ref{eq:coeffCv}) contains the selection rules for the rotational quantum numbers.  The indices $d_{j}$ ($j=0, 1,
2,\ldots$) stand for the states of a given symmetry, open to dissociation at the current energy.  $\alpha$ denotes the eigenchannels
built through the {\it diagonalization} of the reaction matrix \Kmat\, and $-\tan(\eta_{\alpha}^{\Lambda})/\pi$,
$U_{lv,\alpha}^{\Lambda}$ are its eigenvalues and the components of its eigenvectors respectively.

The projection coefficients shown in Eqs.\ (\ref{eq:coeffCv}) and (\ref{eq:coeffCd}) include the two types of couplings controlling the
process: the {\it electronic} coupling, expressed by the elements of the matrices \Umat\ and \etamat, and the {\it non-adiabatic}
coupling between the ionization channels, expressed by the matrix elements involving the quantum defect $\mu_{l}^{\Lambda}(R)$. This
latter interaction is favored by the variation of the quantum defect with the internuclear distance $R$.

\subsection{Construction of the generalized scattering matrix \Xmat}

The matrices \Cmat\ and \Smat\ with the elements given by Eqs.\ (\ref{eq:coeffCv}) and (\ref{eq:coeffCd}) are the building blocks of
the `generalized' scattering matrix \Xmat:
\begin{equation}
\label{eq:Xmatrix}\Xmat =\frac{\Cmat+i\Smat}{\Cmat-i\Smat}.
\end{equation}
It involves all the channels, open (``$o$'') and closed (``$c$''). Although, technically speaking, the open channels only are relevant
for a complete collision event, the participation of the closed channels may influence strongly the cross section, as shown below.

The {\Xmat} matrix relies on 4 block sub-matrices,
\begin{equation}
\label{eq:blocks}
\Xmat =
\left(
  \begin{array}{cc}
    \Xmat_{oo} & \Xmat_{oc} \\
    \Xmat_{co} & \Xmat_{cc} \\
  \end{array}
\right),
\end{equation}
\noindent
where ``$o$'' and ``$c$'' label the lines and columns corresponding to \emph{open} and \emph{closed} channels, respectively.

\subsection{Elimination of closed channels}

The building of the X matrix is performed independently on any account of the asymptotic behaviour of the different channel
wavefunctions.  Eventually, imposing physical boundary conditions leads to the `physical' scattering matrix, restricted to the {\it
  open} channels~\cite{seaton83}:
\begin{equation}
\label{eq:elimination}\SSmat=\Xmat_{oo}-\Xmat_{oc}\frac{1}{\Xmat_{cc}-\exp({
    -i 2 \pi} \numat)} \Xmat_{co}.
\end{equation} 
It is obtained from the sub-matrices of \Xmat\ appearing in Eq.\ (\ref{eq:blocks}) and on a further diagonal matrix \numat\ formed with
the effective quantum numbers ${\nu}_{N^{+}v^{+}}=[2(E_{N^{+}v^{+}}-E)]^{-1/2}$ (in atomic units) associated with each vibrational
threshold $E_{N^{+}v^{+}}$ of the ion situated {\it above} the current energy $E$ (and consequently labelling a \emph{closed} channel).

\subsection{Cross section evaluation}

For a molecular ion initially in the level $N_{i}^{+}v_{i}^{+}$ and recombining with an electron of kinetic (collision) energy
$\varepsilon$, the cross section of capture into {\it all} the dissociative states $d_{j}$ of the same symmetry is given by
\begin{equation}
\label{eq:cs-partial}\sigma _{\text{diss} \leftarrow
  N_{i}^{+}v_{i}^{+}}^{N,\text{\,sym}}=\frac{\pi }{4\varepsilon}\frac{2N+1}{2N_{i}^{+}+1}\rho^{\text{sym}}\sum_{l,\Lambda,j}
|S^{{N\Lambda}}_{d_{j},l N_{i}^{+}v^{+}_{i}}|^{2}.
\end{equation}
On the other hand, the cross section for a ro-vibrational transition to the final level $N_{f}^{+}v_{f}^{+}$, giving collisional
(de\nobreakdash-)excitation, is
\begin{eqnarray}
  \label{eq:partcs} \sigma _{N_{f}^{+}v_{f}^{+} \leftarrow
    N_{i}^{+}v_{i}^{+}}^{N,\text{\,sym}} & = & \frac{\pi }{4\varepsilon
  }\frac{2N+1}{2N_{i}^{+}+1}\rho^{\text{sym}} \nonumber \\
  & & \times 
  \sum_{l,l',\Lambda,j} \left\vert
    S_{N_{f}^{+}v_{f}^{+}l',N_{i}^{+}v_{i}^{+}l}^{N\Lambda}\right\vert
  ^{2}
\end{eqnarray}
Here $\rho^{\text{sym}}$ is the ratio between the multiplicities of the neutral and the target ion. After performing the MQDT
calculation for all accessible total rotational quantum numbers $N$ and for all relevant symmetries, one has to add up the
corresponding cross sections in order to obtain the global cross section for dissociative recombination or ro-vibrational
(de\nobreakdash-)excitation as a function of the electron collision energy $\varepsilon$.

\section{\label{sec_assignment} Computation of cross sections and resonance assignment}

\subsection{Computations}\label{sec_computations}

Since the lowest $^1\Sigma_{g}^{+}$ doubly excited state has by far the most favorable crossing with the ion curve for collisions
taking place at low energy, our analysis has focused on its contribution to the DR cross section.  To improve the accuracy of the cross
sections, we have used for this symmetry the adjusted quasi-diabatic molecular data -- electronic potential curves and couplings -- as
described in Ref.\ \cite{mot08}.  In particular, the quantum defect functions were obtained by the procedure of \textcite{rj87} using
the large set of adiabatic potential curves from \textcite{dsc98}.  With these data, we have performed MQDT-based
\cite{ifs-a18,annick80,valcu} computations of the DR cross section for vibrationally relaxed HD$^{+}$ on its first 13 rotational levels
($N_{i}^{+}=0$--12, $v_{i}^{+}=0$), a range which is by far sufficient to satisfactorily model an equilibrium rotational distribution
corresponding to temperatures below 1000\,K.

For each initial rotational state $N_{i}^{+}$ of the ground vibrational level $v_{i}^{+}=0$, the cross section
$\sigma_{\text{diss}\leftarrow N_{i}^{+}}$ has been obtained by summing up the contributions $\sigma_{\text{diss}\leftarrow
  N_{i}^{+}}^{N,\text{\,sym}}$ over all accessible values of $N$ and all symmetries.  Once the total cross-sections were computed, we
have calculated Boltzmann averages for different rotational temperatures between 100 and 1000~K.  This output has then been used for a
further average, consisting of the convolution with the anisotropic Maxwell distribution of the relative velocities between electrons
and cations in storage ring conditions.  This yields the energy-resolved rate coefficient $\alpha_\text{DR}(E_d)$ described by Eq.\
(\ref{eq:ratecoefficient}) suitable for comparison with the experiment, where appropriate electron temperatures $T_\perp$ and $T_\|$
must be set to define the collision energy distribution $\rho(\varepsilon,E_d)$.  For the calculations presented here we have chosen
$k_BT_\perp=0.5$\,meV and $k_BT_\|=0.02$\,meV.  Regarding the rotational temperature assumed in the Boltzmann average to determine the
rotational level populations, we will assume the value of 300\,K considered as a reasonable estimate of the experimental conditions
(see Sec.\ \ref{sec_expresults}).

The high experimental resolution and the finite number of rotational levels giving significant contributions to the rate coefficient
$\alpha_\text{DR}(E_d)$ suggest to investigate the origin of individual structures in the theoretical cross section and their
correspondence to the energy dependence of the experimental cross section.  Such structure can in particular be of resonant character,
representing electronically bound Rydberg states in the various {\em ionization} channels as defined above, which are associated to
well-defined quantum numbers $N^+,v^+$ of the ion core in the `A-region'.  Single rotational states $N_{i}^{+}$ of the incident ion are
discussed separately before the average over the rotational level populations is added as a final step.  The choice of the electronic
orbital momentum $l$ of the Rydberg resonances corresponds to that applied earlier \cite{ifs-a18}, where it is restricted to $l=0$ and
$2$ ($s$ and $d$ states, only).  The resulting angular momenta $N$ that contribute to the sum in Eq.\ (\ref{eq:cs-partial}) as well as
the channel quantum numbers $N^+$ are those given in Table~1 of Ref.\ \cite{ifs-a18}.

\subsection{Resonance assignment}\label{sec_assign}

The assignment procedure has been carried out for each of the first six rotational states $N_{i}^{+}$ of the ground vibrational state
of HD$^{+}$($^{2}\Sigma_{g}^{+}$), that is, $N_{i}^{+}=0$--5.  As suggested by Eq.\ (\ref{eq:ratecoefficient}), the energy dependent DR
cross section $\sigma_\text{DR}(\varepsilon)=\sigma_{\text{diss} \leftarrow N_{i}^{+}v_{i}^{+}}^{N,\text{\,sym}}(\varepsilon)$ is
represented in the form of $v\sigma_\text{DR}(\varepsilon)$, which is denoted as the ``local'' rate coefficient.  The step to the
experimental energy-resolved rate coefficient then consists only in the convolution with $\rho(\varepsilon,E_d)$ to obtain
$\alpha_\text{DR}(E_d)$.  For simplicity, the same energy scale (denoted as ``electron collision energy'') is used in the figures both
for the collision energy $\varepsilon$ itself and for the detuning energy $E_d$ scanned in the experiment as a proxy for $\varepsilon$;
moreover, the convolved energy-resolved rate coefficient $\alpha_\text{DR}(E_d)$ is represented on the same vertical scale as the
unconvolved local rate coefficient $v\sigma_\text{DR}(\varepsilon)$.  The fully assigned local rate coefficients and the results of the
convolution are presented in Fig.\ \ref{cap:npi1} and \ref{cap:npi2} for HD$^{+}$($^{2}\Sigma_{g}^{+},v_{i}^{+}=0)$ in the initial
ro-vibrational states $N_{i}^{+}=1$ and 2, respectively, which have the highest relative populations at 300\,K (see Table
\ref{tab:rot}).  The energy range of 0--24\,meV is chosen in order to focus on the resonances that determine the DR rate coefficient
under cold plasma conditions up to $\sim$100\,K.

\subsubsection{Results for the most populated initial rotational states}\label{sec_state_results}

\begin{figure}[tb]
  \center\includegraphics[width=8.5cm]{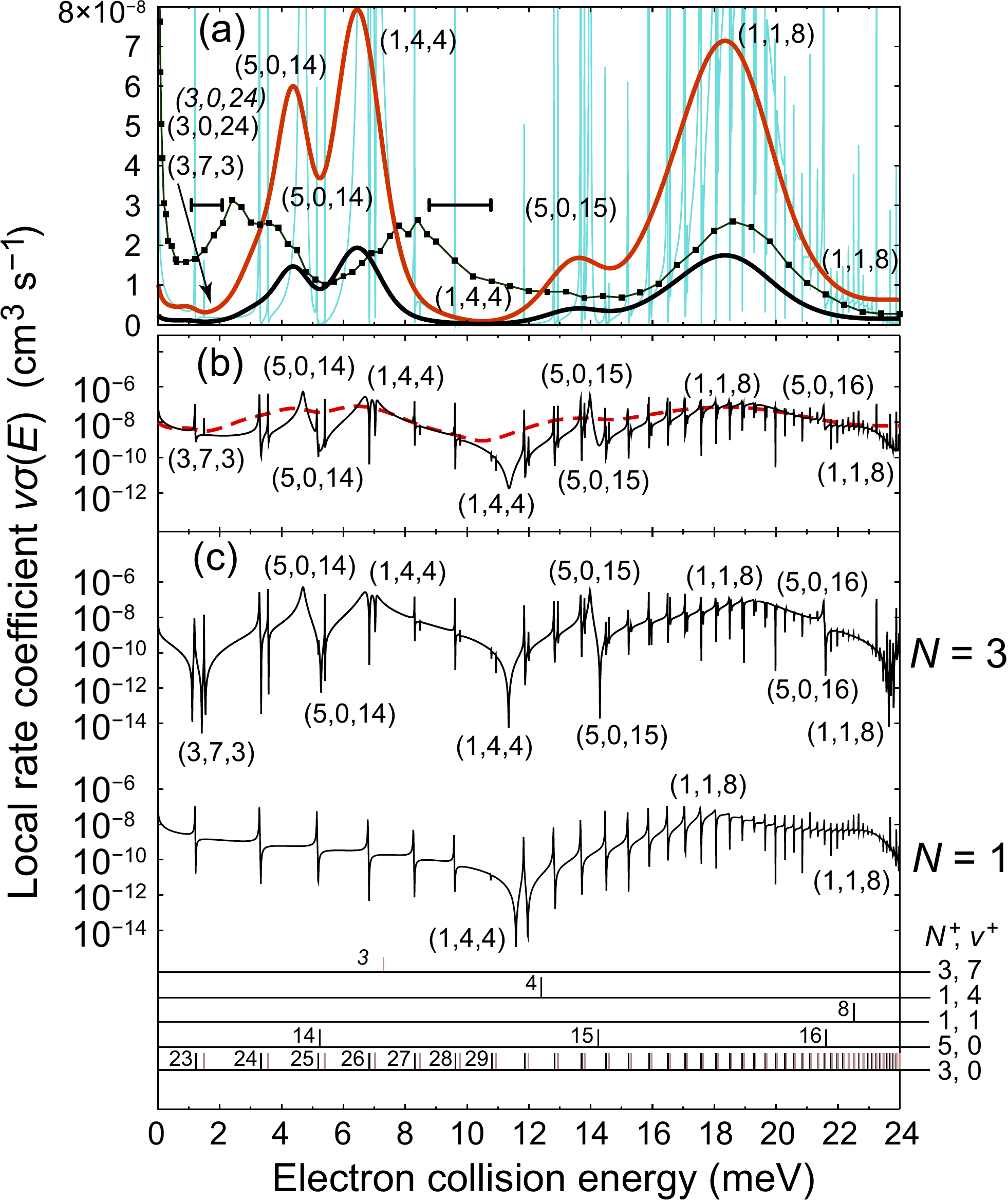}
  \caption{\label{cap:npi1} (Color online) Calculated and assigned MQDT rate coefficient for the DR of
    HD$^{+}$($^{2}\Sigma_{g}^{+},v_{i}^{+}=0$) in the initial state $ N_{i}^{+}=1$.  (a) Convoluted rate coefficient
    $\alpha_\text{DR}(E_d)$ as calculated (upper thick curve) and multiplied by the expected relative Boltzmann population in $
    N_{i}^{+}=1$ (lower thick curve) compared to the measured data (dots and thin line).  Horizontal bars: FWHM experimental collision
    energy spread (see Sec.\ \ref{sec_expresults}).  (b) Local rate coefficient $v\sigma_\text{DR}(\varepsilon)$ summed over the
    contributions of all total angular momenta $N$ (full line) and its convolution (dashed line); the same
    $v\sigma_\text{DR}(\varepsilon)$ is also shown in (a) by a blue (light gray) curve.  (c) Contributions to
    $v\sigma_\text{DR}(\varepsilon)$ from $N=1$--$3$.  At the bottom the energies $\varepsilon$ corresponding to Rydberg states $n$ of
    the relevant channels with core quantum numbers $N^+,v^+$ are indicated by black vertical bars for $d$ states and red (gray) bars
    for $s$ states (numbers for $n$) with the energies from Eq.\ (\ref{eq:ener_cons}) and (\ref{eq:ener_cons_avg}).  Assigned resonance
    minima and maxima are labeled by ($N^+,v^+,n$). Normal type is used for $d$ and italic type for $s$ Rydberg states. }
\end{figure}

\begin{figure}[tb]
  \center\includegraphics[width=8.5cm]{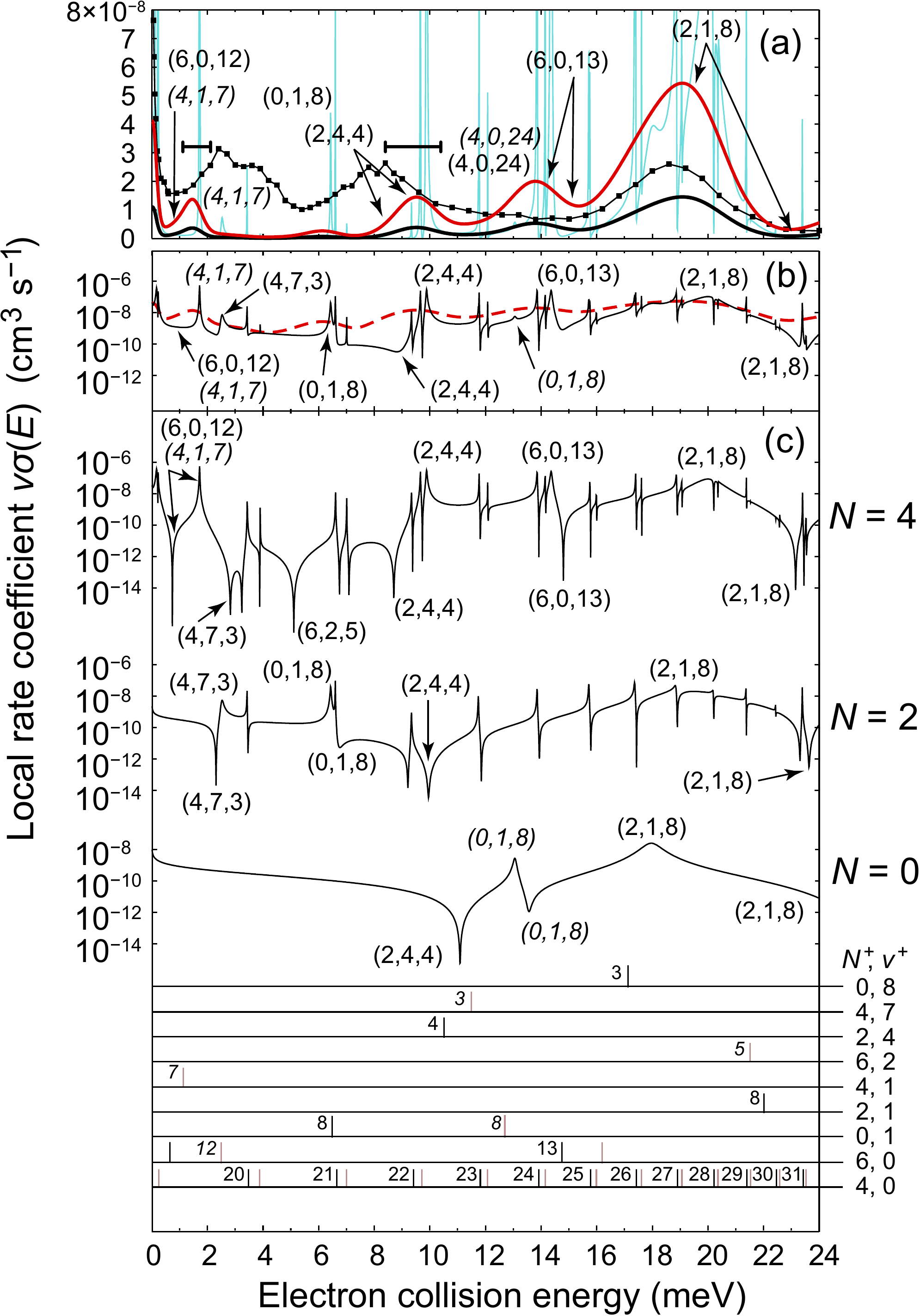}
  \caption{\label{cap:npi2} (Color online) Calculated and assigned MQDT rate coefficient for the DR of
    HD$^{+}$($^{2}\Sigma_{g}^{+},v_{i}^{+}=0$) in the initial state $N_{i}^{+}=2$. See caption of Fig.\ \ref{cap:npi1} for the
    notation.}
\end{figure}

In Fig.\ \ref{cap:npi1} the calculated energy-resolved DR rate coefficient is presented for $N_{i}^{+}=1$.  The result
$\alpha_\text{DR}(E_d)$ of the convolution in Fig.\ \ref{cap:npi1}(a) shows a number of maxima and minima whose assignment has been
obtained by considering the unconvolved local rate coefficient $v\sigma_\text{DR}(\varepsilon)$ in Fig.\ \ref{cap:npi1}(b) and, at the
most elementary level shown in Fig.\ \ref{cap:npi1}(c), the contributions to it arising from the accessible states of total angular
momentum $N$.  Details of the resonance assignment will be discussed in Sec.\ \ref{sec_assign_proc}.  It is guided by considering the
Rydberg series corresponding to various relevant ionic-core excitations as indicated by the prediction bars in the lower part of the
figure.  Purely rotational excitations from $N_{i}^{+}=1$ to $N^+=3,5$ lead to high-$n$ Rydberg resonances in the near-threshold
electronic continuum.  Only small contributions to $\alpha_\text{DR}(E_d)$ result from $N^+=3$ ($n\geq23$), but significant effects
ensue for $N^+=5$, $n=14,15$ even after the convolution.  Significant peaks and dips in the rate coefficient are also caused by higher
ro-vibrational core excitations, which lead to Rydberg resonances with lower principal quantum number $n$.  These structures are
considerably shifted in energy from their positions predicted by the Rydberg formula and need more detailed verifications in order to
identify their origin (see Sec.\ \ref{sec_assign_proc}).  Important structures can be assigned to the resonances $8d$ and $4d$ in the
excitation channels $v^+=1, N^+=1$ and $v^+=4, N^+=1$, respectively, whose energies are well known \cite{tanabe95} to lie in the
electronic continuum relevant for the low-energy DR of HD$^+$.  Note that these cases do not involve any change in the rotation, as
$N^+=N_{i}^{+}$.  Moreover, the $3d$ resonance in the even higher $v^+=7, N^+=3$ excitation channel, energetically predicted below the
ionization threshold for this channel, is found to cause a strong dip in the rate coefficient at only $\sim$2\,meV collision energy for
this incident rotational level.

The energy-resolved DR rate coefficient for $N_{i}^{+}=2$, and the contributions to it of the individual total angular momenta $N$, are
shown in Fig.\ \ref{cap:npi2}.  Considering the core excitations indicated in the lower part of the plot, again some resonances from
purely rotational excitation remain visible even after the convolution over the electron energy distribution.  Thus, the $13d$ Rydberg
level of the excitation channel $N_{i}^{+}=2 \rightarrow N^+=6$ leads to a clear structure around 16\,meV; in addition, the $19s$ and
$19d$ Rydberg levels of the channel $N_{i}^{+}=2 \rightarrow N^+=4$ lie very close to the HD$^+(N_{i}^{+}=2)$ ionization threshold and
likely cause the calculated $\alpha_\text{DR}(E_d)$ to rise by a factor of $>$4 in the range of $E_d<0.5$\,meV.  The dominant
structures in the predicted $\alpha_\text{DR}(E_d)$ are again the $8d$ and $4d$ resonances in the channels $v^+=1$ and $v^+=4$,
respectively, with $N^+=N_{i}^{+}=2$.  The $n=3$ states of the channels $v^+=7, N^+=4$ and $v^+=8, N^+=0$ predicted in this energy
region probably lead to an additional broad modulation of the calculated local rate coefficient.

The comparisons to the experiment in Fig.\ \ref{cap:npi1}(a) and Fig.\ \ref{cap:npi2}(a) show that the calculations for only the two
most populated initial rotational states already well reproduce the observed strong structure near 20\,meV, which coincides with the
predicted $8d, v^+=1$ peaks for both $N_{i}^{+}$.  The calculated positions of these peaks are fairly independent of $N_{i}^{+}$ as the
related resonant capture processes do not involve a change in the core rotation and $n$ is sufficiently high, so that variations of the
quantum defect with $N^{+}$ have only a small effect.  This is different for most other resonances.  The calculated rate coefficient
after summation over $N_{i}^{+}$ will be presented in Sec.\ \ref{sec_results_sum}, where also the main resonant contributions for the
other initial rotational states will be labeled.

\subsubsection{Procedure and verification of resonance assigment}\label{sec_assign_proc}

The resonance assignment demonstrated in the previous section is the result of a two-step strategy, based on the {\em prediction} of
the resonance energies and a {\em validation} of the resonant effect onto the local rate coefficient.  These two steps are implemented
as follows.

\begin{figure}[bt]
  \center\includegraphics[width=8.5cm]{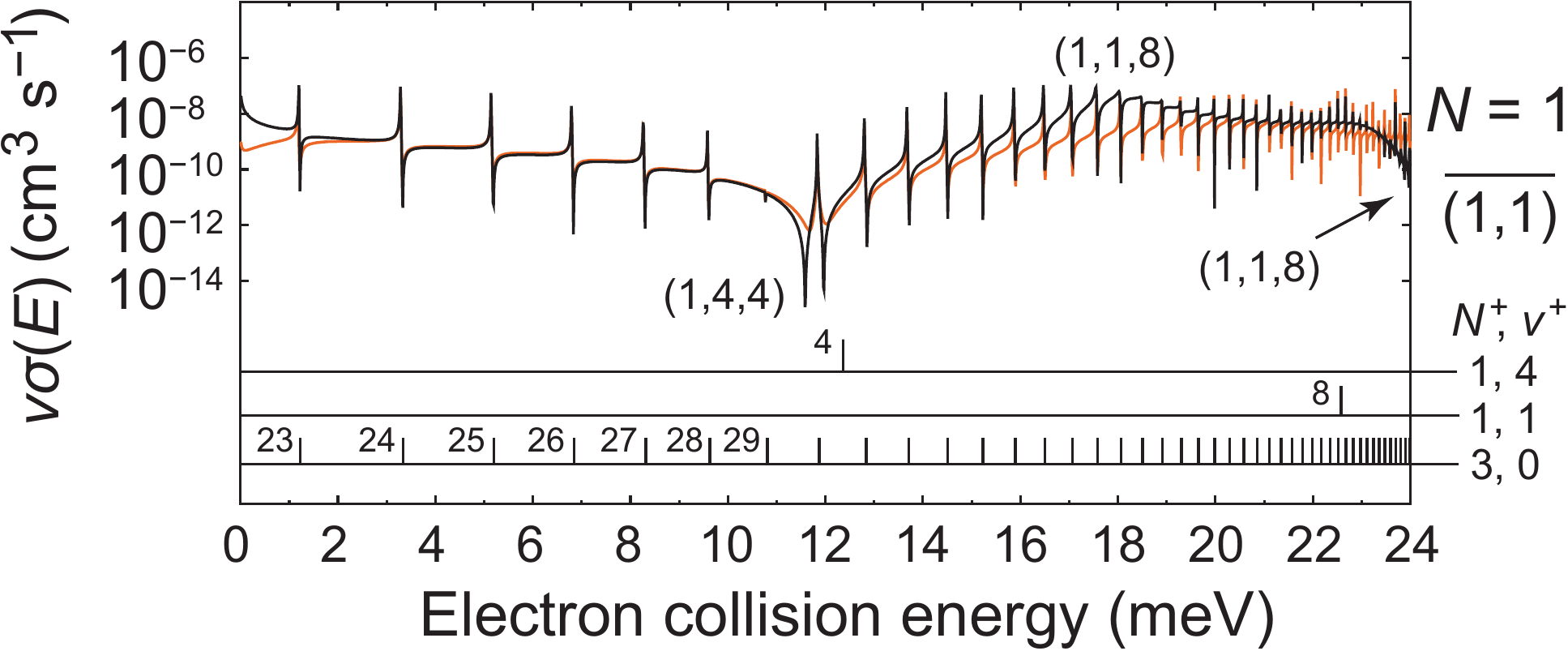}
  \caption[]{\label{cap:demo1} (Color online) Calculated $N=1$ contribution to the MQDT local rate coefficient
    $v\sigma_\text{DR}(\varepsilon)$ for the DR of HD$^{+}$($^{2}\Sigma_{g}^{+},v_{i}^{+}=0$) in $ N_{i}^{+}=1$ and illustration of the
    validation for the assigment of low-$n$ resonances.  Black line: full result for $N=1$; red (gray) line: result obtained on
    eliminating the excitation channel $(N^+,v^+) = (1,1)$ as symbolically indicated in the right margin.  The resonances are labeled
    using the notation of Fig.\ \ref{cap:npi1}.}
\end{figure}

\begin{figure}[bt]
  \center\includegraphics[width=8.5cm]{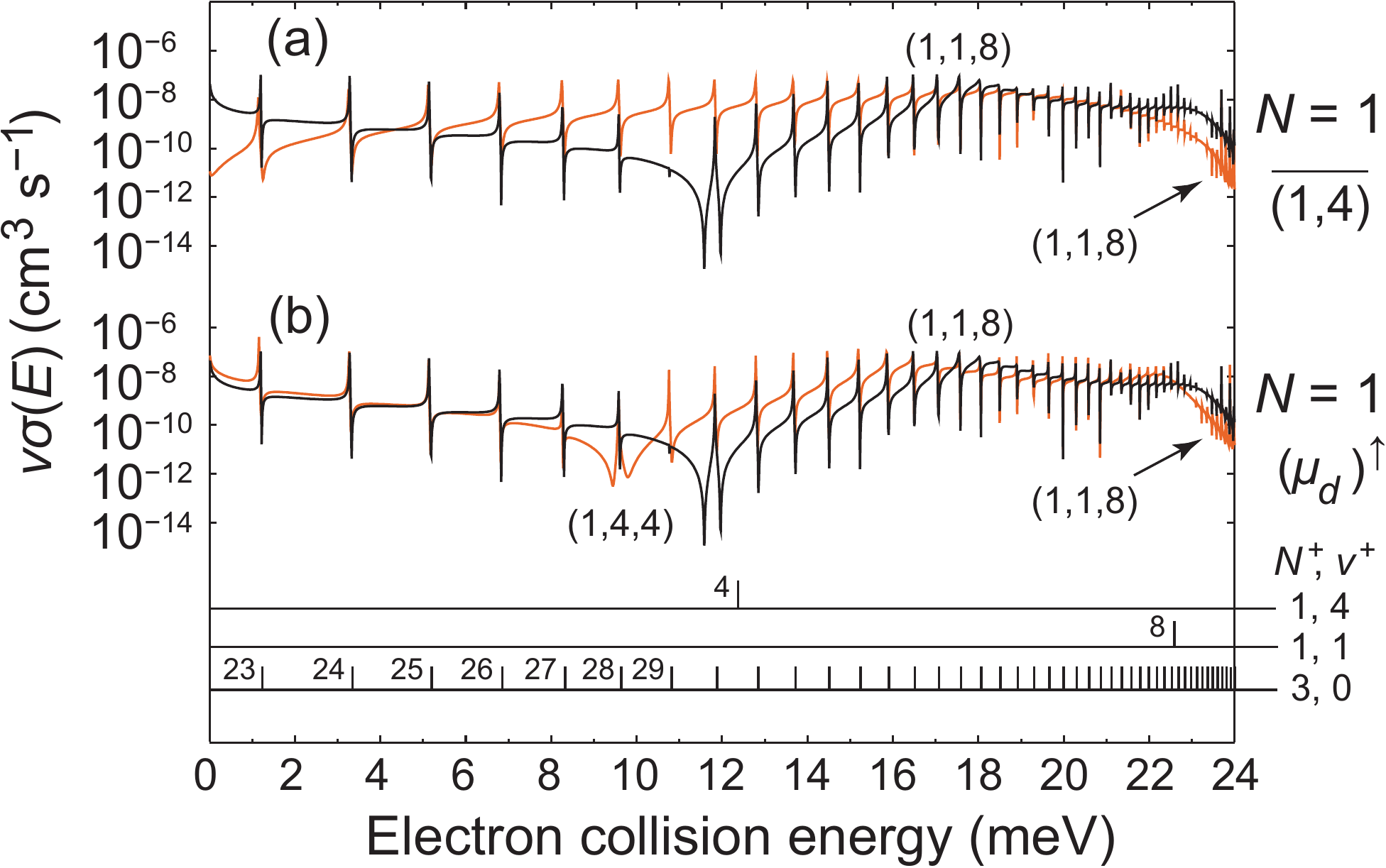}
  \caption{\label{cap:demo2} (Color online) Same rate coefficient as in Fig.\ \ref{cap:demo1} (black line) and validation tests, shown
    as red (gray) lines, by (a) eliminating the $(1,4)$ excitation channel and (b) slightly enlarging the quantum defect of the $d$
    Rydberg levels. Resonance labels use the notation of Fig.\ \ref{cap:npi1}.}
\end{figure}

(i) \emph{Prediction}.
Considering a channel with the core quantum numbers $N^+$ and $v^+$, the conservation of the total energy in the resonant electron
capture process into a Rydberg state with principal quantum number $n$ and orbital angular momentum $l$ can be approximated by
\begin{equation}
  \label{eq:ener_cons} \varepsilon + E\left(
    N_{i}^{+},v_{i}^{+}\right)=E\left(
    N^{+},v^{+}\right)-\frac{\mathcal{R}}{(n-\bar{\mu}_{N^+ v^+ l})^2} .
\end{equation}
Here $\mathcal{R}$ is the Rydberg energy constant, $\bar{\mu}_{N^+ v^+ l}$ the suitably averaged quantum defect of the relevant channel
(see below), and $\varepsilon$ the energy of the incident electron associated with the resonance $nl$ of channel $N^+,v^+$.  These
energies are used to create the Rydberg series prediction bars in the figures of this Section.

The quantum defect can in lowest approximation, appropriate for low $v^+$ and high $n$, be calculated at the equilibrium internuclear
distance $R$ of the ground-state molecular ion.  However, for lower $n$ states, which become relevant at higher $v^+$, the
anharmonicity of the HD$^+$ potential curve together with the $R$-dependence of the quantum defects leads to imprecise predictions of
the Rydberg resonance energies.  Hence, we obtain the quantum defects for a particular excited core state $N^+,v^+$ as the average
\begin{equation}
\label{eq:ener_cons_avg} 
\bar{\mu}_{N^+ v^+ l}  =  \langle \chi_{N^{+}v^{+}}| \mu_{l}^{\Lambda=0}(R)|\chi_{N^+ v^+}\rangle.
\end{equation}
The actual structure of the local rate coefficient due to the Rydberg resonance is still shifted from the prediction of
Eq.~(\ref{eq:ener_cons}) through the perturbation induced by the dissociative state, which is properly taken into account by the MQDT
formalism.  Hence, in a number of cases structures cannot be uniquely identified using the prediction bars alone, and the assignment to
specific excitation channels and Rydberg resonances requires a validation procedure.

(ii) \emph{Validation}.
For a unique identification of the excitation threshold $N^{+},v^{+}$, we eliminate the respective resonant contributions in the
cross-section calculation.  For this purpose, the matrix elements representing the ro-vibronic couplings for this specific threshold
are set to zero in Eq.\ (\ref{eq:coeffCv}) for both ${\cal C}_{lN^{+}v^{+}, \Lambda \alpha}$ and ${\cal S}_{lN^{+}v^{+},\Lambda
  \alpha}$.  In most cases the comparison between the cross section obtained in these trials and the original calculated cross section
enables us to identify the structures due to the considered threshold.  Similarly, the contributions from different resonance orbital
angular momenta $l$ are distinguished by temporarily setting the corresponding quantum defect function $\mu_{l}^{\Lambda}(R)$ to zero
in Eq.\ (\ref{eq:coeffCv}).

A further tool for resonance identification is to probe the sensitivity of the energy at which a structure occurs to slight changes
$\Delta \bar{\mu}_{N^+ v^+ l}$ of the quantum defect, which lead to corresponding changes of the resonance energy according to
\begin{equation}
  \label{eq:distortion} \Delta \varepsilon = - \frac{2\mathcal{R}}{(n-\bar{\mu}_{N^+ v^+ l})^3}\Delta \bar{\mu}_{N^+ v^+ l}.
\end{equation}
Low-lying mono-excited Rydberg states are strongly coupled with the dissociation continuum, which results in large width of the
resonance profile.  Furthermore, the inclusion of the vibronic coupling via the frame transformation induces asymmetric resonance
profiles.  Thus, the assignment of such states using only predicted resonance positions is normally not possible, and the detailed
validation steps discussed here are needed.

\begin{figure}[bt]
  \center\includegraphics[width=8.5cm]{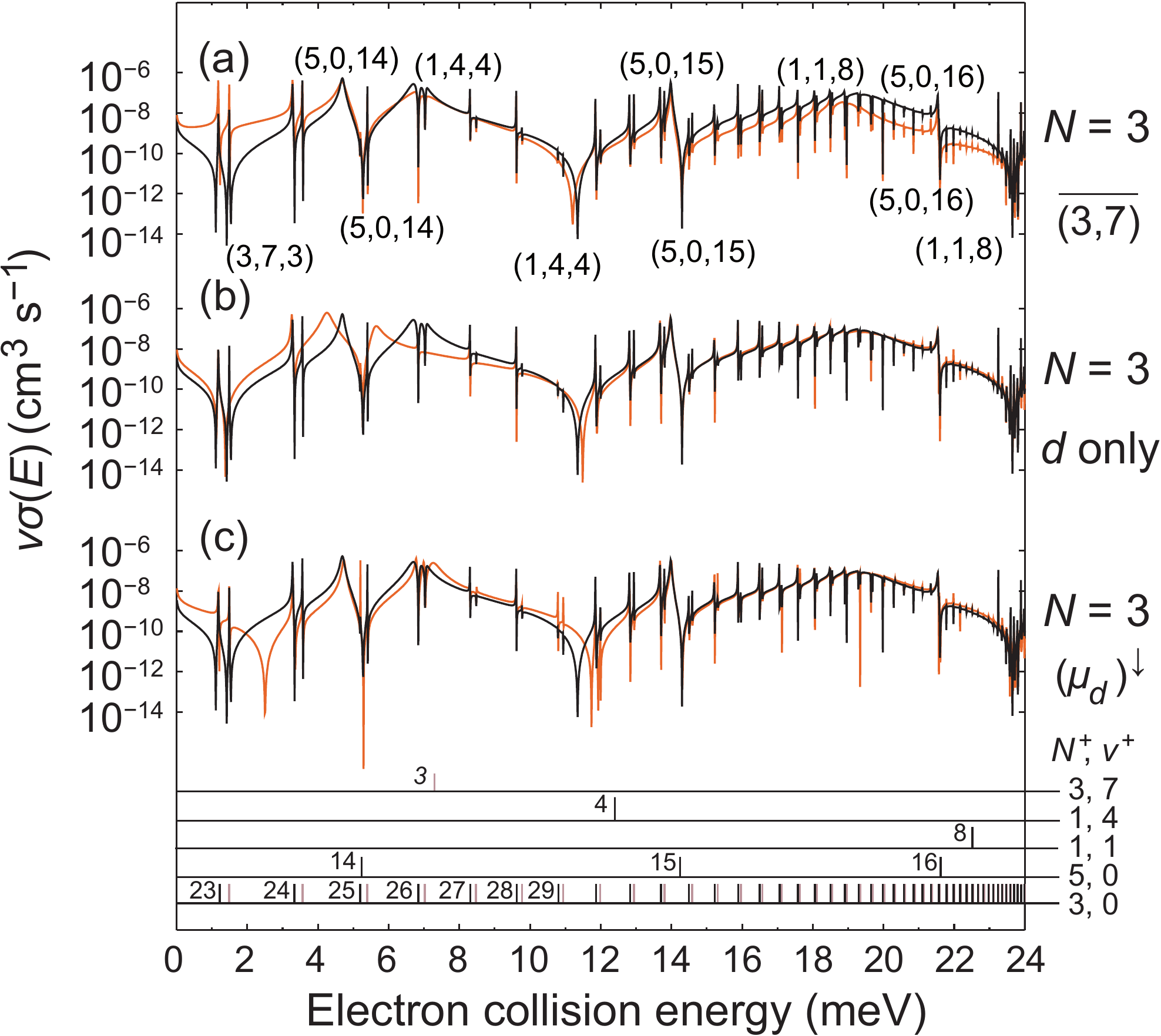}
  \caption{\label{cap:demo3} (Color online) Same rate coefficient as in Fig.\ \ref{cap:demo1}, but for $N=3$ (black line) and
    performing validation tests, shown as red (gray) lines, by (a) eliminating the $(3,7)$ excitation channel, (b) eliminating the
    resonant capture into $s$ Rydberg levels, and (c) reducing the quantum defect of the $d$ Rydberg levels.  Resonance labels use the
    notation of Fig.\ \ref{cap:npi1}. }
\end{figure}

\begin{figure}[bt]
  \center\includegraphics[width=8.5cm]{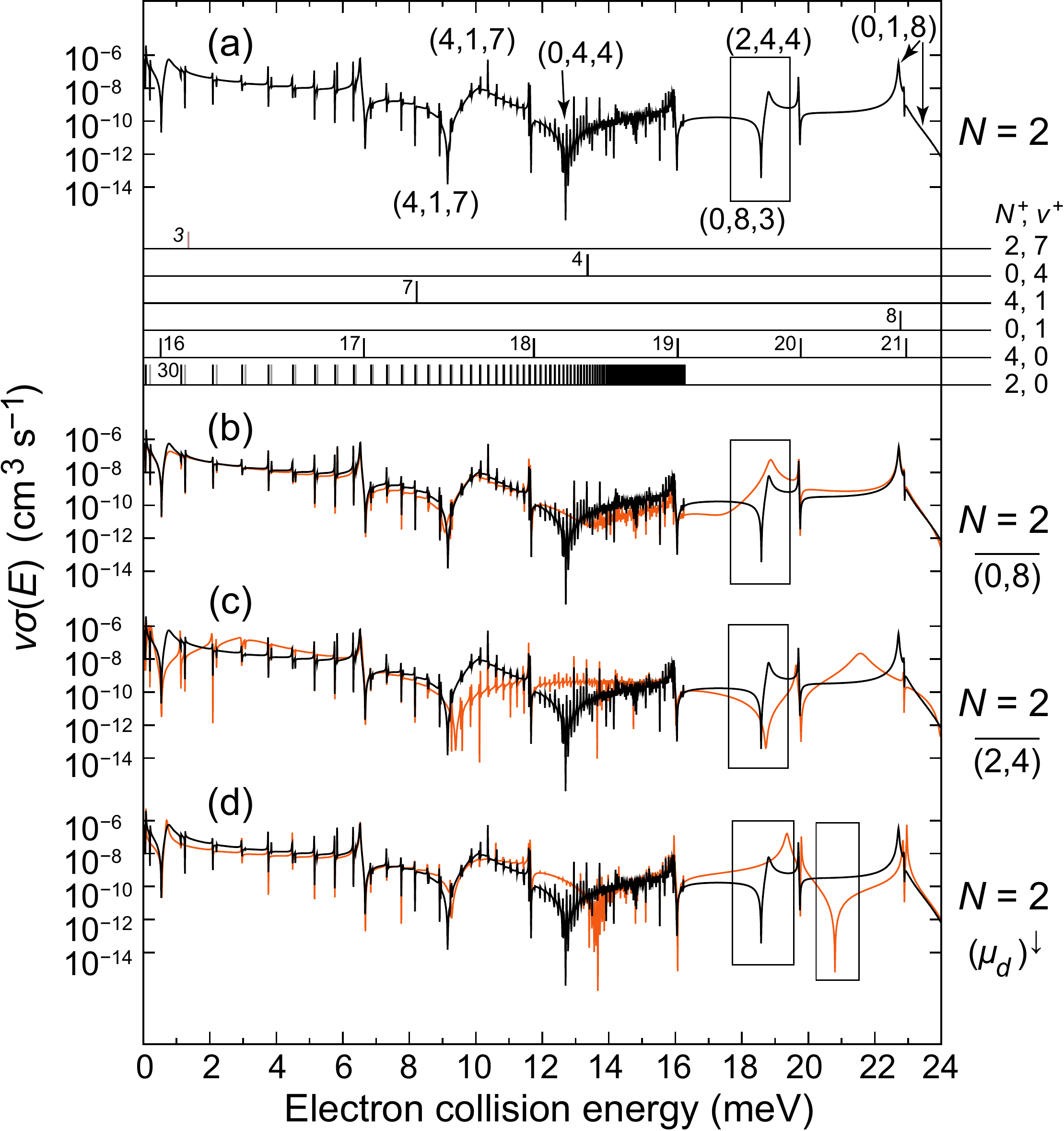}
  \caption{\label{cap:demo4} (Color online) Calculated $N=2$ contribution to the MQDT local rate coefficient for DR of
    HD$^+$($^{2}\Sigma_{g}^{+},v_{i}^{+}=0,N^+_i=0$) ions. (a) Result with resonance labels using the notation of Fig.\ \ref{cap:npi1}.
    The lower traces show the validation procedure of the resonance assignment in particular for the minimum and the maximum marked by
    the rectangle, which turn out to arise from Rydberg states of different series.  Modified rate coefficients are shown by red (gray)
    lines for elimiation of the excitation channels $(N^+,v^+) = (0,8)$ (b) and $(2,4)$ (c) as well as for a reduction of the $d$
    quantum defect (d).  The displaced rectangle in (d) marks the shifted $(0,8)3d$ minimum. }
\end{figure}

The validation of the peak and valley positions for the $8d$ resonance in the channel $(N^+,v^+) = (1,1)$ for $N_i^+=1$ is demonstrated
in Fig.\ \ref{cap:demo1}.  The threshold elimination procedure is used on the $(1,1)$ excitation channel and leads to a disappearence
of the presumed $(1,1)8d$ minima and maxima.  The corresponding validation of the $(1,4)4d$ minimum for the same initial state is
demonstrated in Fig.\ \ref{cap:demo2}(a) regarding the associated excitation channel, while Fig.\ \ref{cap:demo2}(b) shows the effect
of a shift of the quantum defect for the $d$ Rydberg states, which leads to downward displacements of both the presumed $(1,4)4d$ and
$(1,1)8d$ structures, with a larger shift observed as expected for the structure assigned to $4d$.

All three steps were combined to identify the resonance $(3,7)3d$ in the $N=3$ contribution to the $N_i^+=1$ rate coefficient, as shown
in Fig.\ \ref{cap:demo3}.  Near 1.2 meV, the MQDT local rate coefficient shows more mimima than energetically predicted at these
energies (only two minima should occur along the $(3,0)$ Rydberg series from $23d$ and $23s$).  The third minimum could arise from the
nearby $(3,7)3s$ resonance or possibly the $(3,7)3d$ resonance occuring at $\varepsilon<0$.  On removal of all $s$ resonances in Fig.\
\ref{cap:demo3}(b), only the narrow structure close to the predicted $(3,0)23s$ energy disappears. The deepest minimum near 1.2 meV,
occurring between $(3,0)23d$ and $(3,0)23s$, clearly has $d$ character as it remains when the $s$ contribution is removed.  Also, on
slightly lowering the $d$ quantum defect in Fig.\ \ref{cap:demo3}(c), this minimum moves upward by a larger amount than the nearby
$(1,4)4d$, identifying it as $(3,7)3d$.

As a final example for the validation procedure, the identification of resonance maxima and minima is demonstrated in Fig.\
\ref{cap:demo4} for the MQDT local rate coefficient for the DR of HD$^+$ ions in the rotational ground state ($N_i^+=0$).  In addition
to the prediction bars shown, the $(0,8)3d$ and $(2,4)4d$ states are calculated from Eqs.\ (\ref{eq:ener_cons}) and
(\ref{eq:ener_cons_avg}) to lie out of scale, at $\varepsilon=-4.49$\,meV and $\varepsilon=24.96$\,meV, respectively.  Elimination of
the $(0,8)$ excitation channel in Fig.\ \ref{cap:demo4}(b) leads to disappearence of the minimum near 19\,meV, while the elimination of
the $(2,4)$ excitation channel in Fig.\ \ref{cap:demo4}(c) suppresses the nearby maximum.  On a slight decrease of the $d$ quantum
defect in Fig.\ \ref{cap:demo4}(d), both the minimum and the maximum are upshifted, revealing their $d$ character; as expected because
of the lower principal quantum number involved, the minimum assigned to $(0,8)3d$ shifts more strongly than the maximum assigned to
$(2,4)4d$.

\begin{figure}[t]
  \center\includegraphics[width=8.5cm]{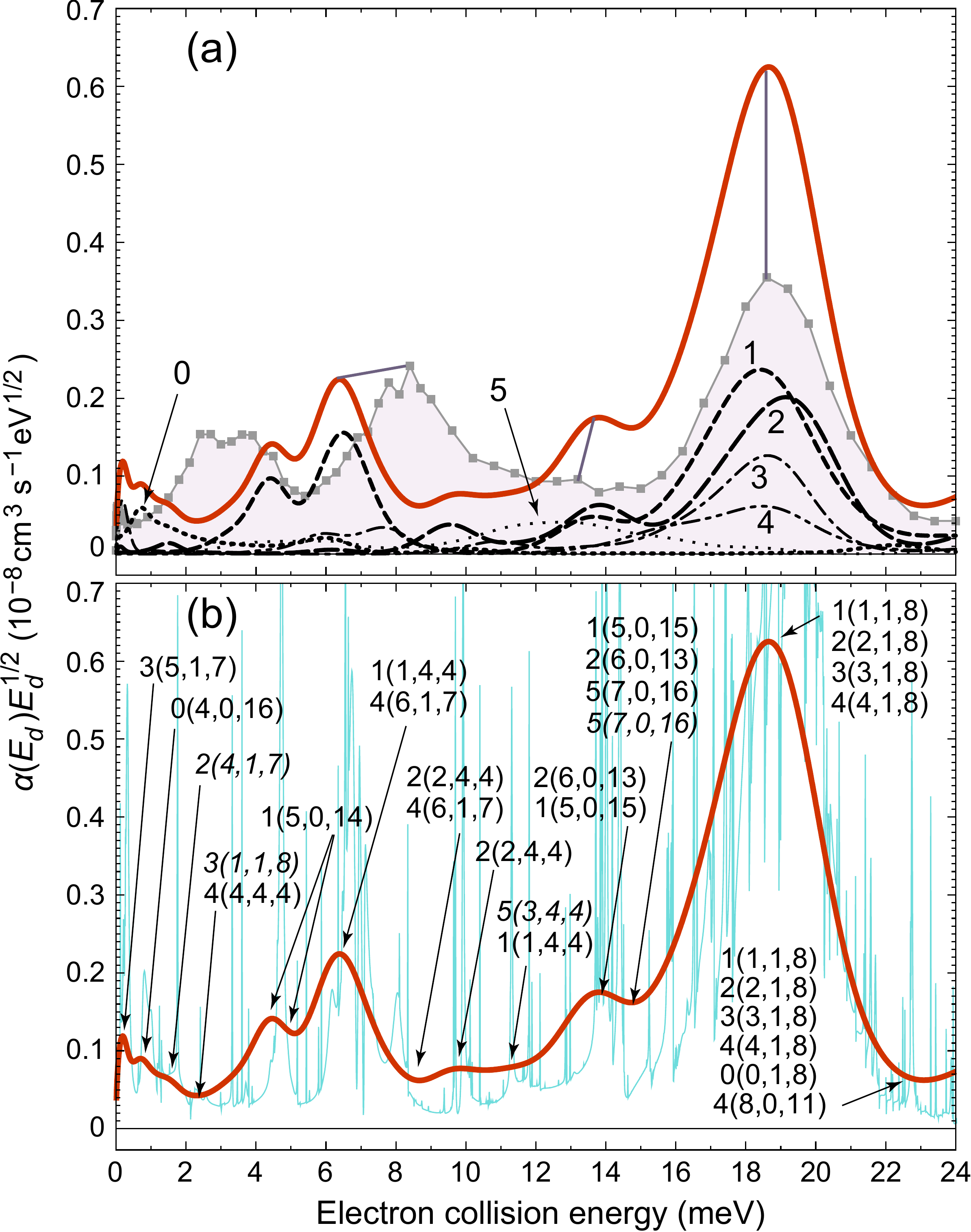}
  \caption{\label{cap:contrib} (Color online) Reduced rate coefficient for the DR of HD$^{+}$($^{2}\Sigma_{g}^{+},v_{i}^{+}=0$). Theory
    after the convolution of Eq.\ (\ref{eq:ratecoefficient}) with $kT_\perp=0.5$~meV and $kT_\|=0.02$~meV.  (a) Contributions of the
    six lowest rotational levels, weighted by the thermal Boltzmann populations at 300\,K (broken lines with numerical labels for
    $N_{i}^+$), total MQDT rate coefficient (thick full curve), and experimental DR rate coefficient (grey dots and thin line).
    Correspondence between experimental and theoretical peaks is indicated by gray connecting lines.  (b) Assignment of the dominant
    contributions to the peaks in the $N_{i}^{+}$-summed MQDT rate coefficient, using the notation $N_{i}^{+}(N^+,v^+,n)$ with normal
    type for $d$ and italic type for $s$ Rydberg states.  Blue (light gray) curve: calculated total rate coefficient
    $v\sigma_\text{DR}(\varepsilon)\varepsilon^{1/2}$ before convolution.}
\end{figure}

\begin{figure*}[bt]
  \center\includegraphics[width=14cm]{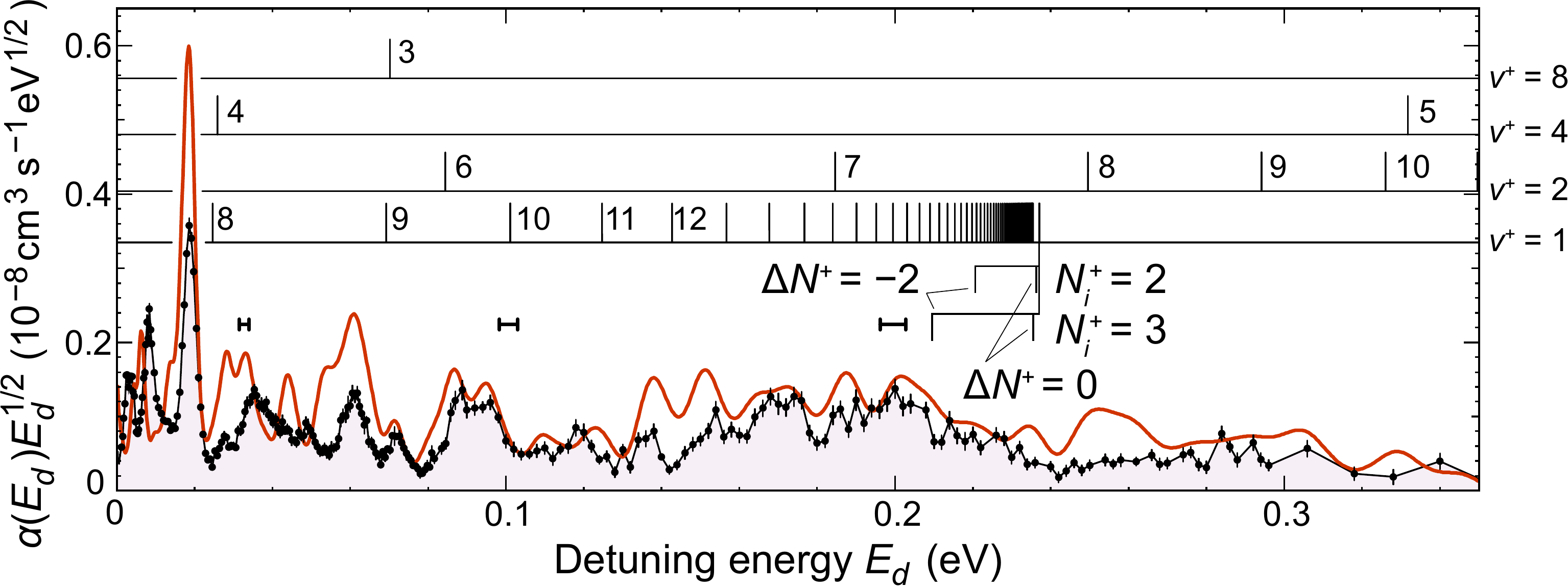}
  \caption{\label{cap:theovsexp} {(Color online) Reduced DR rate coefficients for the DR of HD$^{+}$($^{2}\Sigma_{g}^{+},v_{i}^{+}=0$)
      from experiment (black dots and thin line) and MQDT theory (thick solid curve) after collision energy convolution and initial
      state averaging as in Fig.\ \ref{cap:contrib}.  Prediction bars: Rydberg resonance energies (labeled by principal quantum numbers
      $n$; zero quantum defect) below the energy thresholds for vibrational excitation to levels $v^+$ as indicated, assuming $\Delta
      N^+ = 0$ in the resonance formation.  Near the $v^+=1$ threshold, its energetic shifts due to molecular rotation are indicated
      for initial states $N_i^+ = 2,3$ and the cases $\Delta N^+ = 0$ and $\Delta N^+ = -2$.  Horizontal bars: FWHM experimental
      collision energy spread in the respective energy regions (see Sec.\ \ref{sec_expresults}).}}
\end{figure*}

\section{\label{sec_comparison} Detailed comparison between storage-ring and MQDT results}
\label{sec_results_sum}

For the comparison between experimental and theoretical results, we choose the reduced DR rate coefficient
$\alpha_\text{DR}(E_d)E_d^{1/2}$ which emphasizes the resonant variations.  The Boltzmann-weighted contributions of the individual
HD$^+$ rotational states as well as the total MQDT rate coefficient obtained by summing these contributions are shown in Fig.\
\ref{cap:contrib}(a), while the main resonances assigned to the structures in the total MQDT rate coefficient are indicated in Fig.\
\ref{cap:contrib}(b).  For the shape of the theoretical rate coefficient, the $N_{i}^{+}=1$ contribution turns out to have the largest
influence.  In fact, the predicted double structure between 4 and 7\,meV resulting from the $(1,4)4d$ and $(5,0)14d$ resonances for
this $N_{i}^{+}$ remains visible also in the sum over all initial rotational states.  HD$^+$ ions in $N_{i}^{+}=1$ are also predicted
to yield the largest contribution to the peak near 19\,meV through the $(1,1)8d$ resonance.  This resonance is further emphasized by
other resonances of the same Rydberg level and with $N^{+}=N_{i}^{+}$, which occur at essentially the same $\varepsilon$ for
$N_{i}^{+}=1$--4.  Near 14\,meV the maxima and minima caused by the Rydberg levels $(5,0)15d$ of $N_{i}^{+}=1$ and $(6,0)13d$ of
$N_{i}^{+}=2$ add up.  Theory also predicts two structures at very low energy $\varepsilon\lesssim1$\,meV: a peak due to combined
ro-vibrational excitation of $N_{i}^{+}=3$ ions, caused by a $7d$ Rydberg level, and a rather strong $16d$ Rydberg resonance through
purely rotational excitation of $N_{i}^{+}=0$ ions to $N^{+}=4$.  The latter prediction would significantly influence the DR of HD$^+$
in a low-temperature environment, where $N_{i}^{+}=0$ will be much more important than in the present case.  [Note that also the $n=29$
Rydberg levels of the $N^{+}=2$ excitation channel occur in very close vicinity of the ionization threshold ($\varepsilon=0$) for these
ions, as shown in Fig.\ \ref{cap:demo4}, but the MQDT calculation predicts lower contributions to the DR rate for these states.]
Altogether, while the strongest resonances in the calculated rate coefficient are due to combined ro-vibrational excitation, also
purely rotational excitation and capture in Rydberg levels with $n=13$--16 are found to give significant contributions.

In comparison to the experiment, a clear correspondence is observed between the calculated peaks of $N_{i}^{+}=1$--4 near 19\,meV and
the dominant experimental structure found at this energy.  At lower energies, less good agreement is found; in particular, the
prominent experimental peaks at 2--4\,meV (with a sub-structure) and at 8\,meV 
are not correctly reproduced by the theory.  Considering the trend of low-$n$ resonances to acquire a large width, it appears not
unreasonable to suggest a correspondence between the strong theoretical $(1,4)4d$ resonance for $N_{i}^{+}=1$ at 6~meV with the
experimental structure observed 2\,meV higher in energy.  It can also be rationalized that for the low $n$ involved in such resonances,
both energetic position and width will sensitively depend on the underlying quantum defect of this channel and on its residual $n$
dependence.  On the other hand, this assumes that the lower experimental peak (or unresolved peaks) are caused by levels not properly
accounted for in the calculation.  It must be noted that for a predicted high-$n$ resonance, such as $(5,0)14d$ for $N_{i}^{+}=1$ near
5 meV, variations of the quantum defect will not significantly change the resonance position on the considered energy scale (even
though the resonance strength and the exact position of the associated minima and maxima will be affected).  Hence, the 2--4\,meV
experimental peak can be plausibly assumed to be partly caused by the $(5,0)14d$, $N_{i}^{+}=1$ resonance, while other contributions to
it are obviously missed by the calculation.

The theoretical results obtained in the same approach over a broader energy range, including the excitation limit to the first HD$^+$
vibrational level at 0.237\,eV, are shown in Fig.\ \ref{cap:theovsexp} and compared to the experimental data, using again the reduced
rate coefficient $\alpha_\text{DR}(E_d)E_d^{1/2}$.  It should be noted that the thermal rate coefficient of $\alpha_\text{th} =
(9\pm2)\times10^{-9}$ cm$^3$\,s$^{-1} \times (300\,\text{K}/T)^{1/2}$ determined in the limit of low electron temperature by
\textcite{alkhalili03} translates into a constant low-energy limit of $\alpha_\text{DR}(E_d)E_d^{1/2}= (\pi
k_BT)^{1/2}\alpha_\text{th}/2$ that amounts to $(0.13\pm0.03)\times10^{-8}$\,cm$^3$\,s$^{-1}$\,eV$^{1/2}$, in good agreement with the
average value of the reduced rate coefficient observed at $<$0.02\,eV, while the overall trend of $\alpha_\text{DR}(E_d)E_d^{1/2}$ then
shows a lower average up to $\sim$0.2\,eV.  A further decrease occurs as expected at the first vibrational excitation threshold, when
the autoionization into the electronic continuum of HD$^+(v^+=1$) becomes energetically allowed following resonant electron capture
(resonant vibrational excitation).

The energetic thresholds for autoionization plotted in Fig.\ \ref{cap:theovsexp}, which are complemented by the attached Rydberg series
of capture resonances, assume $N_{i}^{+}=0$ incident ions and are slightly shifted by rotational effects.  Thus, for $N_{i}^{+}>0$ and
autoionization into the continuum with $N^{+}=N_{i}^{+}$, the reduction of the vibrational constant by centrifugal effects moves the
threshold positions downward by a few meV, resulting in shifts of $-0.74$ and $-1.48$\,meV for $N_{i}^{+}=2$ and $N_{i}^{+}=3$ incident
HD$^+$ ions, respectively.  The shifts become larger for autoionization into the continuum with $N^{+}=N_{i}^{+}-2$, corresponding to
vibrational excitation connected with rotational de-excitation and are illustrated for the $v^+=1$ threshold and the case $\Delta N^+
=-2$ in Fig.\ \ref{cap:theovsexp}, where they amount to $-16.3$ and $-27.3$ meV, respectively.  Since processes with $\Delta N^+ \neq
0$ are clearly important in the MQDT calculations, as discussed in detail in Sec.\ \ref{sec_assignment}, all resonance positions marked
here (in contrast to Sec.\ \ref{sec_assignment}) can have only indicative character considering the rotational excitation relevant in
the present study.  With this in mind, for Fig.\ \ref{cap:theovsexp} also the quantum defects were neglected in the
Rydberg state positions marked which, however, mainly affects the low-$n$ resonances for $v^+=4$ and 8.

The present MQDT result yields overall a remarkable agreement with both the magnitude and the structure of the observed rate
coefficient.  In fact, up to $\sim$0.2\,eV most prominent structures of the reduced rate coefficient are resonably well reproduced by
the theory with local energy shifts below 5\,meV.  Also an intermittant dip in $\alpha_\text{DR}(E_d)E_d^{1/2}$ between 0.10 and
0.14\,eV is reasonably borne out by the MQDT prediction.  Both in experiment and theory, the drop in the reduced rate coefficient at
$>$0.2\,eV, i.e., already below the rotationless vibrational excitation threshold, is remarkable.  In fact, the onset of this decrease
occurs just at the expected threshold energies of $\Delta N^+ =-2$ processes for $N_{i}^{+}=2$ and 3 indicated in Fig.\
\ref{cap:theovsexp}, which suggests that resonant electron collisions followed by autoionization in this energy region lead to
vibrational excitation together with rotational de-excitation.  The initial rotational levels for which this effect can be expected are
among the most populated ones at 300\,K (see Table \ref{tab:rot}).

\section{\label{sec_conclusions} Discussion and outlook}

At the high energy resolution and theoretical accuracy achieved in the present work, dissociative recombination experiments using a
stored beam of vibrationally relaxed HD$^+$ ions turn out to be a sensitive tool for studying rotational effects in resonant low-energy
collisions with molecular cations.  The energy-resolved rate coefficient reveals a rich structure from ro-vibrationally excited Rydberg
resonances which drive the DR processes observed here as well as the other collision channels of ro-vibrational
(de\nobreakdash-)excitation and elastic scattering.  Extensive MQDT calculations yield a high level of agreement with the observed
structures, which confirms the good quality of the molecular data applied, the efficiency of the theoretical approach for predicting
the low-energy resonant processes, and the assumptions about the rotational level population underlying the interpretation of the
experimental results.

The rotational level population, close to the thermal equilibrium caused by radiative transitions in the blackbody radiation field of
the 300-K (room-temperature) storage-ring and by the inelastic and superelastic electron collisions in the electron cooler, leads to
$\sim$5 similarly occupied HD$^+$ initial levels with excitation energies well above the experimental energy resolution.  The largest
structure, a narrow resonance near 19\,meV, is assigned to the rotationally excited initial states $N_i^+=1$--4, whereas no
contribution from $N_i^+=0$ is predicted for this peak.  The molecular rotation $N^+$ is conserved in the formation of these electron
capture resonances.  The calculation predicts similarly large rate coefficients for processes with $\Delta N^+ =\pm2, \pm4$ which,
however, for the various initial rotational levels are strongly dispersed over the energy scale studied here.  While the experimental
structures are not reproduced in all their detail, the agreement between the measured structures and those in the MQDT rate coefficient
after convolution with the experimental energy distribution is remarkable and by far the best for any molecular ion to-date, especially
at this resolution.  Not only can most observed structures between 5 and 200~meV be individually assigned to calculated ones, but also
the ratio of the calculated and the measured rate coefficients remains within 0.5 to 2.0 over essentially this entire energy range.

The ro-vibrational resonant structure in the low-energy DR rate coefficient of HD$^+$ is among the most pronounced ones observed for
this process on any molecular cation.  Since ro-vibrational resonances of the type considered here occur for a wide range of diatomic
and polyatomic ions, their statistically averaged DR rate coefficients in cold media should be expected to be significantly influenced
by the magnitudes and the temperature dependences of this resonant DR process.  However, most experiments so far observed rather
structureless energy dependences in low-energy DR studies of heavier systems \cite{drbook}.  It is likely that this can be attributed
to the much smaller energetic differences between the rotational levels for these species as compared to the hydrogen cation, which
leads to an even larger number of rotational levels populated at the rotational temperatures of at least 300\,K in the collision-energy
resolved DR measurements so far performed on such systems.  This stands in contrast to the low-temperature conditions of many media in
which DR is important, and much more structure in the energy dependence of the DR rate coefficient may well be revealed also in these
cases with improved control over the rotational populations.  For HD$^+$, the present results predict that the collision energy
dependence of the DR rate coefficient for rotationally cold ($N_i^+=0$) ions will be dramatically different from that observed here and
in previous experiments with similar, near room-temperature, rotational populations. The development of a cryogenic electrostatic ion
storage ring suitable for electron--ion collision experiments with merged beams is in progress \cite{zajfman-jpcs05,lange-rsi10} and
will make energy-resolved DR studies possible under conditions where rotational excitation by the thermal blackbody radiation is
largely eliminated and the parent ions are rotationally cold.

On the theoretical side, the present work demonstrates the power of the MQDT method for predicting low-energy collision cross sections
between electrons and molecular cations determined by numerous Rydberg capture resonances, including a detailed account of rotational
effects.  The magnitude and energetic dependence of the DR rate coefficient can be well understood on the basis of these calculations
after averaging over the electron energy distribution and the initial HD$^+$ rotational distribution.  Methods for identifying the
individual resonant contributions in the theoretical cross section even in cases of overlapping and strongly broadened resonances have
been implemented.  This will be of great help in future efforts to further improve the quantum defect functions and their
implementation in the resonant cross-section calculation.  Full experimental control over the initial rotational excitation of the
HD$^+$ ions, in connection with MQDT calculations using the procedures presented here, can be expected to yield a complete
understanding of the state-to-state resonant collision mechanism for this benchmark reaction also in the low-temperature limit which is
not yet finally explored.

%
\begin{acknowledgments}
  The authors would like to thank Ch.~Jungen, H.~Takagi, C.~H.~Greene, V. Kokoouline, and S.~L.~Gubermann for numerous discussions and
  the TSR and accelerator teams at the Max-Planck Institute for Nuclear Physics for their support during the beam times.  We
  acknowledge scientific and financial support from the European Spatial Agency/ESTEC 21790/08/NL/HE, the International Atomic Energy
  Agency (CRP ``Light Element Atom, Molecule and Radical Behaviour in the Divertor and Edge Plasma Regions''), the French Research
  Federation for Fusion Studies (CEA, EFDA, EURATOM), the French ANR project `SUMOSTAI', the CNRS/INSU programme `Physique et Chimie du
  Milieu Interstellaire', the projects Triangle de la Physique/PEPS `Physique th\'{e}orique et ses interfaces', CPER
  Haute-Normandie/CNRT `Energie, Electronique, Mat\'{e}riaux', and PPF/CORIA-LOMC `Energie-Environnement', and the agencies IEFE
  Rouen-Le Havre, and Conseil R\'egional de la Haute Normandie.  IFS thanks the Laboratoire Aim\'e--Cotton and the Max-Planck Institute
  for Nuclear Physics for their hospitality.  XU is senior research associate of the Belgian Fund for Scientific Research (FNRS).  The
  experimental work was supported by the Israel Science Foundation, by the German Israel Foundation for Scientific Research (GIF) under
  contracts I-707-55.7/2001 and I-900-231.7/2005, and by the European Project ITS LEIF (HRPI-CT-2005-026015).  DS acknowledges support
  from the Weizmann Institute through the Joseph Meyerhoff program.  The support by the Max-Planck Society is gratefully acknowledged.
\end{acknowledgments}
%

%

%

%


\end{document}